\documentclass{aastex}          
\usepackage{spr-astr-addons}    

\newcommand{\HII}{\hbox{H\,{\sc ii}}}
\newcommand{\HI}{\hbox{H\,{\sc i}}}

\def\p0{\phantom{0}}

\begin{document}

%
\title{An analysis of the FIR/RADIO Continuum Correlation\\ in the Small Magellanic Cloud}

\shorttitle{FIR/RC correlation in the SMC}
\shortauthors{Leverenz}

\author{Howard Leverenz\altaffilmark{1}}
\affil{James Cook University, Townsville, QLD 7000, Australia}
\email{hleverenz@sbcglobal.net} 
\and
\author{Miroslav D. Filipovi\'c\altaffilmark{1,2}}
\affil{University of Western Sydney, Locked Bag 1797, Penrith South DC, NSW 1797, Australia}
\affil{James Cook University, Townsville, QLD 7000, Australia}
\email{m.filipovic@uws.edu.au} 


\begin{abstract}

The local correlation between far-infrared (FIR) emission and radio-continuum (RC) emission for the Small Magellanic Cloud (SMC) is investigated over scales from 3~kpc to 0.01~kpc. Here, we report good FIR/RC correlation down to $\sim$15~pc. The reciprocal slope of the FIR/RC emission correlation (RC/FIR) in the SMC is shown to be greatest in the most active star forming regions with a power law slope of $\sim$1.14 indicating that the RC emission increases faster than the FIR emission. The slope of the other regions and the SMC are much flatter and in the range of 0.63--0.85. The slopes tend to follow the thermal fractions of the regions which range from 0.5 to 0.95.  The thermal fraction of the RC emission alone can provide the expected FIR/RC correlation. The results are consistent with a common source for ultraviolet (UV) photons heating dust and Cosmic Ray electrons  (${\rm{CR}}{{\rm{e}}^ - }$s) diffusing away from the star forming regions. Since the ${\rm{CR}}{{\rm{e}}^ - }$s appear to escape the SMC so readily, the results here may not provide support for coupling between the local gas density and the magnetic field intensity.

\textbf{Accepted for publication in Astrophysics \& Space Science}

\end{abstract}

\keywords{galaxies: individual: Small Magellanic Cloud - Magellanic Clouds - infrared: galaxies - radio-continuum: galaxies.}

\section{Introduction}\label{s:intro}

One of the most puzzling and robust relationships in extragalactic astronomy is the virtually ubiquitous correlation between far-infrared (FIR) and radio-continuum (RC) measurements of star-forming galaxies \citep{yun01} (hereafter referred to as Y01). The essentially linear relationship extends for over five orders of magnitude \citep{cond92} exhibiting the amazingly small scatter of $~\sim 0.2~dex$  (z~dex~$=~{10^z})$. Virtually all star forming galaxies are included: normal barred and unbarred spiral galaxies, irregular and dwarf galaxies, and Seyferts including radio-quiet quasars. This correlation has been found to apply to galaxies even beyond z=1 \citep{appl04}.

The standard explanation for these observations is found in models that use the conversion of energy resulting from the formation of young massive stars ($ > 20~M_ \odot $). FIR emission is posited to be the result of dust heated by UV radiation from these massive young stars. These stars also provide energy for the thermal radio emission found in star forming regions. Massive stars quickly evolve to the terminus of their life cycle and explode into supernovae. The supernova remnants (SNRs) expand rapidly into the local interstellar medium (ISM) and the shock wave provides the mechanism for accelerating Cosmic Ray electrons (${\rm{CR}}{{\rm{e}}^ - }$s) \citep{held09,schu09}. Those electrons interact with the galactic magnetic field and are thus responsible for the non-thermal (synchrotron) component of the RC emission \citep{cond92}. They also interact with particles in the ISM and contribute to the thermal (free-free) component of the RC emission along with UV radiation from hot massive stars.

This simple picture cannot explain the small dispersion in the FIR/RC ratio measured among the diversity of galaxies which span the observed ranges of star formation rates (SFRs), magnetic field strengths, metallicity, dust grain chemistry, and ISM mass values.

The scale at which models break down provides an important constraint on the physical mechanisms detailed in that model. For galaxy-wide scales some models work well on spiral galaxies. For very small scales, for example, within several kpc of the Sun, the FIR/RC ratios of star forming regions are not consistent with expected FIR/RC values. Instead, the radio emission appears to be primarily thermal \citep{hasl87}.

\cite{hugh06} examined the Large Magellanic Cloud (LMC) for 60~$\mu$m FIR / 21~cm RC correlations on spatial scales from 1.5 to 0.05~kpc and found a tight correlation on spatial scales above $\sim$ 50~pc.

The SMC can be described as a gas-rich late-type dwarf galaxy \citep{bola07}. It has a gas-to-dust ratio 17 times higher than the Milky Way Galaxy \citep{koor84}. It is a member of the local group and is classified as irregular (ImIV-V) \citep{sand94}. It may also be a satellite galaxy of the Milky Way \citep{west97}. The SMC is centered at $60.6 \pm 3.8$~kpc \citep{hild05} from the Galaxy. The spatial scale of the SMC is 0.3pc/$''$. The line of sight to the SMC has an extinction of only AV$ \sim$0.2 magnitudes and a reddening of E(B-V)$ \sim$0.04 \citep{bola07,schl98}.

The SMC appears to have been tidally disrupted by a recent close encounter with the LMC \citep{mura80}. There is no trace of any spiral structure in the SMC \citep{sand81} but both a $''$bar-like$''$ structure and a $''$wing$''$ have been reported \citep{west97}. The bar structure is probably the body of the SMC and contains most of its gas and star formation activity. The wing feature continues into the bridge which is an \HI\ region extending from the SMC to the LMC \citep{bola07}. The SMC may be more complex than the LMC considering that various population centroids vary by $\sim 20'$ in R.A. and by $ \sim {1^ \circ }$ of declination \citep{west97}. The SMC has an absolute luminosity of $M_{\rm B} = 2.9m$  \citep{west97}.

The SMC has the lowest metallicity of any gas-rich galaxy in close proximity \citep{bola07}. This makes it one of the best galaxies to provide understanding of processes in very high redshift galaxies. Measurements suggest that the dust grains are mostly silicates rather than carbonaceous grains which distinguish the SMC's ISM from that of the Galaxy \citep{wein01}.

The stars resolved in the SMC are blue and red supergiants with $M_{\rm B}=-7m$. These Population~I stars are accompanied by a resolved region of Population~II stars with $M_{\rm V}=-3m$. \cite{schw93} identified 249 infrared sources in the SMC. The morphology of the Population~II stars has been measured and appears to suggest a spheroidal distribution without the prominent irregular features seen in the younger stars \citep{bola07,cion00,mara01,zari00}. Earlier studies of the motion of \HI\ clouds were interpreted as representing rotation \citep{west97}.  Recent observations, however, find no evidence for rotation in the SMC \citep{piat08}, \citep{cost11}.

In this paper, Section \ref{s:data} discusses the origins and characteristics of the data used in terms of the frequencies of each data set, characteristics of the data, and source instruments. Section \ref{s:Analysis} discusses the analysis methods applied to the data sets.  Section \ref{s:Discussion} is a discussion of results of the analysis and considers some FIR/RC galaxy models. Data processing and analysis for this paper was done with Miriad \citep{saul95}, Karma \citep{gooc96}, Matlab \citep{matl10}, DS9 \citep{joye03} and Ftools \citep{blac95}.

\section{Observational Data}
 \label{s:data}

The data used for this study are summarized in Table~\ref{tbl:list}. The FIR data are from the IRAS \citep{mivi05} and the Spitzer satellites \citep{bola07}. The RC data are from the Australian Telescope Compact Array (ATCA) which consists of five movable 22-m antennae, and the Parkes 64m radio telescopes.  (For reference see the series of papers on radio-continuum studies of the Magellanic Clouds: \cite{hayn91,xu92,klein93,fili95,fili97,fili98,fili98a,fili98b}, \cite{fili02,payn04,fili05,reid06,payn07}, and \citep{craw11,wong11,wong11a,wong12}).

\begin{figure*}
\includegraphics[width=\textwidth,height=\textwidth]{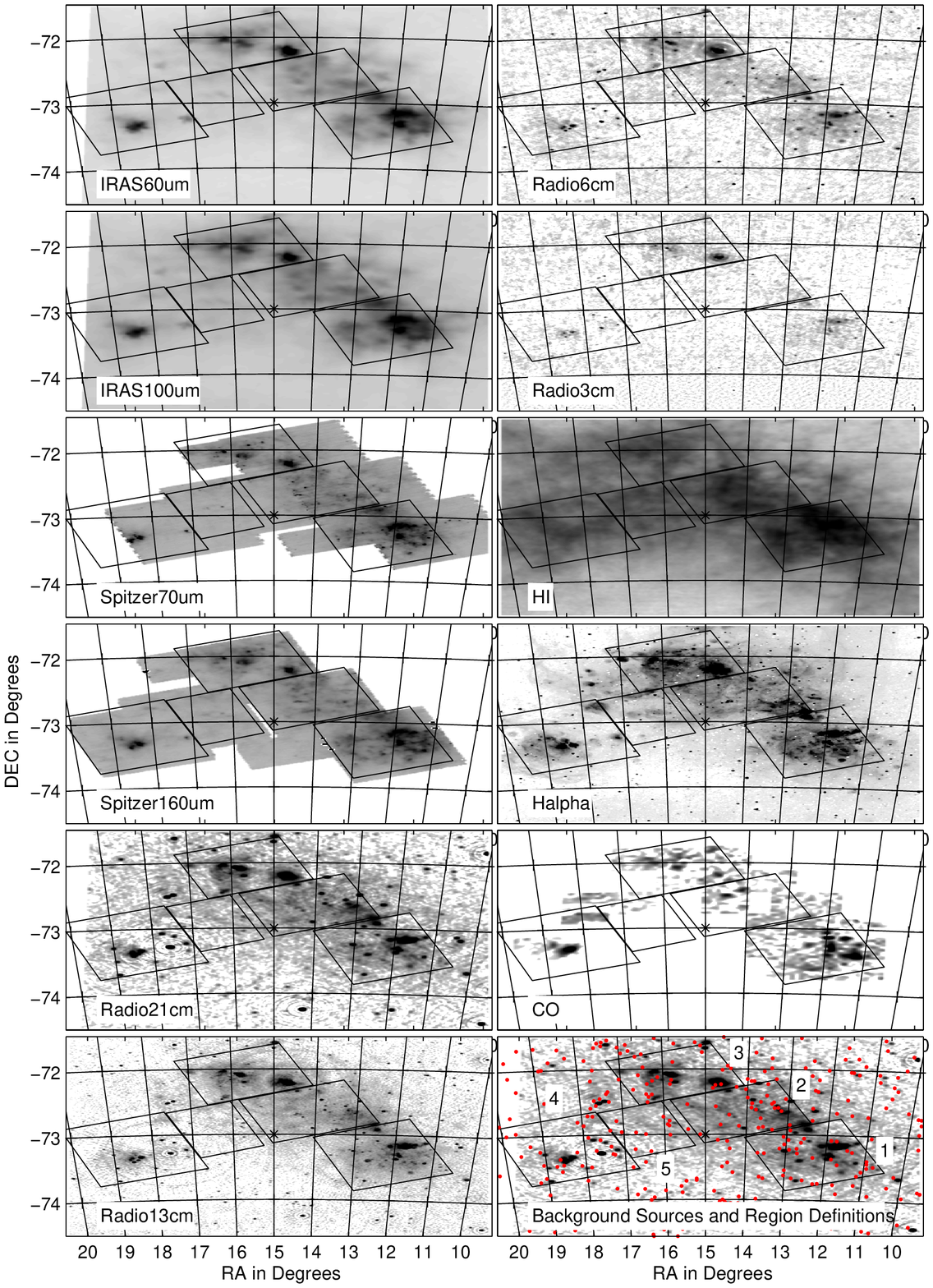}
\caption{All of the data sources used in this paper are shown in this figure. The IRAS FIR data are the improved reprocessed data from the IRAS satellite survey \citep{mivi05}. Spitzer FIR data are from the Spitzer Survey of the Small Magellanic Cloud (S3MC) \citep{bola07}. The $160~\mu m$ Spitzer data provides the outline for the definition of the 5 regions used in this analysis which is shown in the background source panel. The four sources of RC data used in this paper are shown in this figure. \HI\ data from the ATCA and Parkes telescopes are combined to produce the data used here. Details of the data are found in Table \ref{tbl:list}. Data are displayed with North to the top and East to the left and J2000 coordinates in degrees. These plots are smoothed with a 25$''$ Gaussian kernel and scaled by the square root of the intensity.}
\label{fig:data}
\end{figure*}

Also, included are data sets for $\HI$ \citep{stav97,stan99}, H$\alpha$ \citep{smit00}, and CO  \citep{mizu01} detailed below. In order to make comparisons between the data sets, all of the data are regridded to a common pixel size of 5$''$ using a gnomonic projection centered at $ 15^\circ\ $RA and $ -73^\circ\ $DEC (equatorial J2000 coordinates). The whole SMC is defined as the data shown in Figure~\ref{fig:data}. The irregular coverage of the SMC by the Spitzer data suggested that a division of the data sets into several regions would be appropriate. The 160~$\mu$m Spitzer data is used as an approximate template for the division into the five regions. Figure~\ref{fig:data} shows the regional definitions superimposed on all of the data sets as defined by the $160~\mu m$ Spitzer data. It is clear that the bar of the SMC is represented by regions~1-3 and the wing by regions~4 and 5. Table~\ref{tbl:coordinates} lists the coordinates of the boundaries and centers of the regions as well as their areas. Table~\ref{tbl:fluxes} contains the flux densities from each region and from the whole SMC.

%
\begin{table*}
\small
\caption{Data used in this study include IRAS and Spitzer satellite images,
Radio images from ATCA and Parkes telescopes, \HI\ images, H$\alpha$ images from NFCCD/CTIO
and CO $j=(1-0)$ images. } 
\label{tbl:list}
\begin{tabular}{@{}llccl@{}}
\tableline  
Band & Telescope & Resolution & Resolution & Reference \\
 & & (arcsec) & (pc) \\
\tableline
60~$\mu m$ & IRAS & 120 & 35.3 & \citep{mivi05} \\
100~$\mu m$ & IRAS & 120 & 35.3 & \citep{mivi05} \\
70~$\mu m$ & Spitzer & 18 & 5.3 & \citep{bola07} \\
160~$\mu m$ & Spitzer & 40 & 11.8 & \citep{bola07} \\
3~cm RC & ATCA+Parkes & 2 & 0.6 & \citep{dick10} a \\
6~cm RC & ATCA+Parkes & 3 & 0.9 & \citep{dick10} a \\
6~cm RC & Parkes & 162 & 48.6 & \citep{fili97} a \\
13~cm RC & ATCA+Parkes & 60 & 17.6 & \citep{stav97,stan99} a \\
21~cm RC & ATCA+Parkes & 90 & 26.4 & \citep{stav97,fili97} a \\
21~cm RC & Parkes & 1128 & 338 & \citep{fili97} a \\
\HI\ & ATCA+Parkes & 98 & 28.8 & \citep{stav97,stan99} a \\
H$\alpha$ & NFCCD/CTIO & 3 & 0.9 & \citep{smit00} \\
CO & NANTEN 4-m &156 & 45.8 & \citep{mizu01} \\
\tableline 
\end{tabular}
%
\tablenotetext{a}{Also see \cite{hayn91,xu92,klein93,fili95,fili97,fili98,fili98a,fili98b}, \cite{fili02,payn04,fili05,reid06,payn07}, and \cite{craw11,wong11,wong11a,wong12}}
\end{table*}

\begin{table*}
\small
\caption{Coordinates in degrees of the boundaries of the 5~regions as shown in Figure~\ref{fig:data}. Each region is defined by the coordinates of the region's vertices. The centers of each region and the areas are also shown.}
\label{tbl:coordinates}
\begin{tabular}{@{}cccccccccccc@{}}
\tableline  

Vertex & SE &   SW &  NW &  NE & Center &  Area \\
\tableline
Regions & RA/DEC & RA/DEC & RA/DEC & RA/DEC & RA/Dec & $Deg^2$ \\
\tableline
1&12.9528/-73.8571&10.4595/-73.5514&11.7169/-72.7425&14.0390/-73.0425&12.2920/-73.2984&0.6647\\
2&15.0297/-73.1336&12.4418/-72.8108&13.3596/-72.1624&15.8403/-72.4741&14.1679/-72.6452&0.5767\\
3&16.5846/-72.5451&14.0808/-72.2471&14.8865/-71.6050&17.3130/-71.8711&15.7163/-72.0671&0.5679\\
4&19.4525/-73.7627&16.6597/-73.5182&17.6519/-72.6450&20.2119/-72.8758&18.4940/-73.2004&0.7432\\
5&16.8303/-73.3567&15.2447/-73.1735&16.0482/-72.4953&17.6098/-72.6634&16.4332/-72.9222&0.3589\\
SMC&20.5906/-74.4285&9.4094/-74.4285&10.2887/-71.4425&19.7253/-71.4424&15.0000/-73.0000&9.0000\\
\tableline 
\end{tabular}
%
\end{table*}

\begin{table*}
\small
\caption{Regional measured data and derived results for \HI\ in terms of ${M_ \odot }$ and $H_2$ mass from CO measurements.  The percentages refer to the portion of the total summed over all regions for each wavelength shown in this table. The SMC data refers to the whole SMC data set which includes data not contained in the regions. }
\label{tbl:fluxes}
\begin{tabular}{@{}lccccccl@{}}
\tableline  
Data & Region 1 & Region 2 & Region 3 & Region 4 & Region 5 & SMC & Units \\
\tableline  
60 $\mu m$ & 24.9(39\%)&13.1(20\%)&16.0(25\%)&7.33(11\%)&2.88(4\%)&92.6&${10^2}$ Jy \\
100 $\mu m$ & 47.6(37\%)&26.4(21\%)&29.3(23\%)&17.7(14\%)&7.47(6\%)&208.&${10^2}$ Jy \\
70 $\mu m$ & 26.6(34\%)&18.8(24\%)&17.0(22\%)&9.02(12\%)&5.98(8\%)&102.&${10^2}$ Jy \\
160 $\mu m$& 73.9(35\%)&42.2(20\%)&45.7(22\%)&34.3(16\%)&15.2(7\%)&332.&${10^2}$ Jy \\
\HI\ & 2.64(30\%)&1.87(22\%)&1.41(16\%)&1.79(21\%)&0.954(11\%)& 48.2 & ${10^{27}}$ \HI\ /$cm^2$ \\
\HI\ & 4.56(30\%)&3.23(22\%)&2.43(16\%)&3.09(21\%)&1.65(11\%)&27.9& ${10^7}$ \HI\ ${M_ \odot }$ \\
H$\alpha $ &44.1(30\%)&30.2(20\%)&54.7(37\%)&15.6(10\%)&4.78(3\%)&149.& $10^{-11}$ergs/$cm^2$/s\\
CO & 494.(52\%)&121.(13\%)&179.(19\%)&150.(16\%)&4.72(0\%)&82.6& ${10^2}$ K km/s\\
$H_2$ & 71.5(50\%)&13.9(9.7\%)&32.8(23\%)&24.6(17\%)&0.54(0.4\%)&159.&${10^5} {M_ \odot }$ \\
3~cm& 6.28(26\%)&3.80(16\%)&7.61(32\%)&4.49(19\%)&1.57(7\%)&54.8& Jy\\
6~cm& 5.80(29\%)&3.87(19\%)&6.47(32\%)&2.91(15\%)&0.861(4\%)&33.5& Jy\\
13~cm& 7.59(31\%)&5.69(23\%)&7.63(31\%)&2.94(12\%)&0.680(3\%)&35.1& Jy\\
21~cm& 7.55(29\%)&6.46(24\%)&7.59(29\%)&3.69(14\%)&1.17(4\%)&40.0& Jy\\
\tableline 
\end{tabular}
%
\end{table*}


%

\subsection{Infrared Data} \label{s:infrared}

The IRAS data used in this paper are the improved reprocessed data from the IRAS satellite survey \citep{mivi05}. The data are in two bands: $60~\mu m$ and $100~\mu m$ and have been corrected for zodiacal light, calibration, zero levels and striping problems (see Figure~\ref{fig:data}).

The data used from the Spitzer satellite are from the MIPS instrument. They are from the Spitzer Survey of the SMC (S3MC) covering most of the bar and wing \citep{bola07} in two bands, $70~\mu m$ and $160~\mu m$. The data are corrected for zodiacal light, Galactic foreground and Cosmic Infrared background using Leroy's published data \citep{lero07} to adjust the final calibration.

\subsection{Radio-Continuum Data}  \label{s:radiocontinuum}

Data from two radio telescopes are used in this study; the Parkes Radio Telescope and the Australian Telescope Compact Array (ATCA). For high-resolution correlation investigations, the combined ATCA+Parkes data is used and for the low resolution thermal fraction investigations, the Parkes data from \cite{fili97} is used exclusively due to its superior overall flux-density accuracy.

The high resolution RC data at 21~cm and 13~cm are from observations made with the ATCA telescope in 1992 from October $6^{th}$ to the $8^{th}$. The observations were made using five antennae in the 375~m east-west array configuration \citep{wong12,fili05} at 1.42~GHz and at 2.37~GHz. The angular resolution is $ \sim$98$''$ (FWHM) and $ \sim$40$''$ (FWHM). There were 320 pointings used to create a mosaic of data covering an area of $\left( { \sim 20^\circ } \right)^2 $. The $''$short spacing$''$ low spatial frequency data fill is accomplished using Parkes survey data \citep{fili97}.

The 6~cm and 3~cm high resolution continuum data are from \citep{dick10}, at 4.8~GHz and 8.6~GHz, using the EW352 (February 2005) and EW367 (February and March 2006) configurations in 3564 individual pointing positions achieving resolutions of 35$''$ and 22$''$ respectively. The data were reprocessed to include data from a sixth antenna approximately 6~km from the center of the ATCA as well as including the low spatial frequency data from the Parkes survey \citep{hayn91}. The addition of this incomplete, 2$''$-3$''$ interference pattern, provides greater resolution to the image for the evaluation of previously unresolved sources and will be used here to enhance morphological studies using wavelet cross correlations. The additional data does not cover the full extent of the SMC but covers all of the area studied in this paper. See \cite{hayn91,xu92,klein93,fili95,fili97,fili98,fili98a,fili98b}, \cite{fili02,payn04,fili05,reid06,payn07}, and \cite{craw11,wong11,wong11a,wong12}.

\subsection{\HI\ Data}\label{s:HI}

The \HI\ 21~cm data are from observations made with the ATCA telescope in 1992 from October $6^{th}$ to the $14^{th}$ and August $2^{nd}$ and $3^{rd}$. The observations were made using five antennas in the 375~m east-west array configuration \citep{stav95,stav97} at 1.42~GHz. The angular resolution is $ \sim$90$''$ (FWHM). There were 320 pointings used to create a mosaic of data covering an area of $\left( { \sim 20^\circ } \right)^2 $. The $''$short spacing$''$ low spatial frequency data fill is accomplished using Parkes survey data \citep{stan99} from 1996 March $14^{th}$ to $17^{th}$.

The column densities are corrected for self-absorption using the correction factors from \cite{stan99}. The \HI\ mass is calculated for each region and the whole SMC and is shown in Table~\ref{tbl:fluxes}. Results reported here of $2.8 \cdot {10^8}{\text{ }}HI{\text{ }}{M_ \odot }$ from a $3^{\circ}$ $\times$ $3^{\circ}$ reduced data set compare favorably to a published value of $3.8 \cdot {10^8}{\text{ }}HI{\text{ }}{M_ \odot }$ from \cite{stan99} for a $5^{\circ}$ $\times$ $5^{\circ}$ image.

\subsection{H$\alpha$ Data}\label{s:Halpha}

The H$\alpha$ data is from the Magellanic Cloud Emission Line Survey (MCELS) program using the UM/CTIO Curtis Schmidt telescope \citep{smit00}. The central $3.5^\circ\ $$\times$ $4.5^\circ\ $ of the SMC are imaged in overlapping $1^\circ\ $$\times$ $1^\circ\ $ fields with $2048^2$ pixels per field in such a way as to provide at least two samples of each pixel. The H$\alpha$ filter is centered on 500.7~nm with a width of 3~nm. Two continuum channels are also imaged, 685~nm, $\delta=9.5~nm$, and 513~nm, $\delta=3~nm$ for stellar contribution subtraction.

\subsection{$ {}^{12}CO{\text{ }}j = (1 - 0)
$ Data (115.271 GHz)}\label{s:CO}

CO data is from the NANTEN sub-millimeter observatory on Pampa la Bola in the Atacama Desert, Chile \citep{mizu01}. The NANTEN telescope is used at 2.6~mm wavelength and the observations have a resolution of $2.6'$. Regions~1-3 are well covered by this data. Region~4 has coverage primarily around N84 and there is only incidental coverage of Region~5.

The conversion factors for the CO data to H$_2$ mass used in this paper are from \cite{lero11}. Leroy calculated 3 different conversion factors for the SMC for different parts of the SMC which varied by 40\%. (Leroy's West~$\rightarrow$~Region~1; North~$\rightarrow$~Region~3; East~$\rightarrow$~Region~4; Average of Leroy's West and North $\rightarrow$ Region~2 and Region~5). The results of this conversion is expected to reveal only the most dense H$_2$ clouds. The conversion factors applied to the 5~regions and the overall average value applied to the entire SMC are shown in Table~\ref{tbl:coregions}.

\begin{table}
\small
\caption{CO conversion factors to $H_2$ from \cite{lero11} as assigned to the SMC regions and the calculated $H_2$ mass for each region.}
\label{tbl:coregions}
\begin{tabular}{@{}ccc@{}}
\tableline  
Region & Conversion Factor & $H_2$  \\
 &${{M_ \odot }p{c^{ - 2}}{(K{\text{ }}km{\text{ }}{s^{ - 1}})^{ - 1}}}$
 & ${*10^5} {M_ \odot }$ \\
\tableline
1  & 67 & 71.5 \\
2  & 53 & 13.9 \\
3  & 85 & 32.8 \\
4  & 67 & 24.6 \\
5  & 53 & 0.54 \\
SMC & 69 & 159 \\
\tableline 
\end{tabular}
%
\end{table}

\subsection{Background sources}\label{s:BKG}

Since the SMC is essentially transparent to 1.42~GHz RC radiation and it presents a cross section of several deg$^2$, many background sources are expected to be visible in the SMC RC data. These background sources were catalogued by \cite{payn04} \cite{craw11,wong11,wong11a,wong12}. This data consist of 717 RC sources of which 616 were identified as background sources. They are plotted in Figure~\ref{fig:data} along with the numbers assigned to each of the five regions on an image of the 21~cm RC data.

\section{Analysis}
 \label{s:Analysis}

The analysis presented here uses cross correlations between several different wavelengths using traditional pixel-by-pixel correlations as well as wavelet~correlations. The pixel-by-pixel correlation studies provide intensity correlations as a function of position. The wavelet cross correlation technique probes morphological correlations which are sensitive to changes over a range of spatial scales. FIR/RC correlation is studied over the entire SMC as well as the regions defined previously.

All of the correlation analysis with RC data is done excluding data at the coordinates of the background sources. The mask consisted of Gaussian function with a FWHM of $2'$ applied to the coordinates of each background source. The data sets also had a median filter of $15''$~$\times$~$15''$ applied to remove zero pixels and provide a slight smoothing.

\subsection{SMC and Regional FIR/RC ratios}
 \label{s:FIRRC}

\begin{figure*}
\includegraphics[width=\textwidth,height=\textwidth]{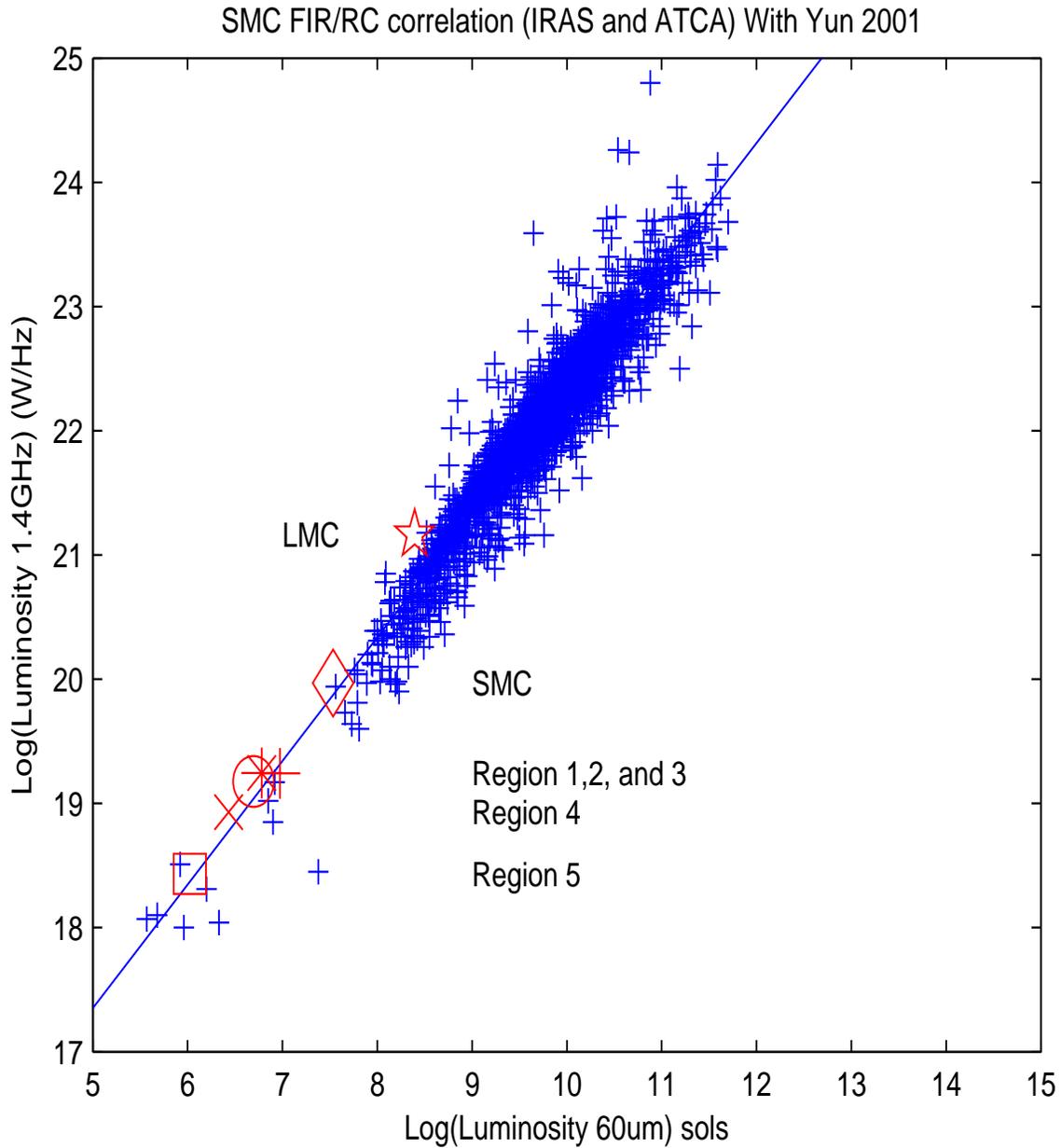}
\caption{FIR/RC correlations for the whole SMC (defined as a $ 2.5^\circ\ $ x $ 2.5^\circ\ $  region centered at RA $ 15^\circ\ $ Dec $ -73^\circ\ $) and the regions defined in this paper compared to the data from \cite{yun01}. Regions 1-5 and whole SMC are plotted and labeled. (1-plus, 2-circle, 3-asterisk, 4-x, 5-square, and SMC-diamond). Additionally, the correlation calculated from data published on the LMC by \cite{hugh06} is plotted using the star marker.}
\label{fig:yuncorr}
\end{figure*}

An examination of the FIR/RC correlation in the SMC compared to measurements from more distant galaxies is shown in Figure~\ref{fig:yuncorr}. This correlation is measured down to scale factors of less than 1~kpc by considering each of the five regions in the SMC (see Figure~\ref{fig:data}). Three of these regions cover the bar structure and two cover the portion of the wing structure that is closest to the bar. The structure of the correlations within the regions is shown in images which display the organization of the FIR and RC emissions giving rise to the measured values.

Using the $60~\mu m$ IRAS IR flux and the 21~cm RC flux, a FIR/RC calculation is made for each of the five SMC regions and the whole SMC (defined as a 3$^\circ\ $$\times$ 3$^\circ\ $region centered at RA~$ 15^\circ\ $ Dec~$ -73^\circ\ $). These results are directly compared to the survey done by Y01. Y01's catalog identified radio counterparts to the IRAS Redshift Survey galaxies with $60~\mu m$ IR fluxes of $ \gtrsim 2~Jy $. The catalog includes 1750 galaxies and contains radio positions and redshifts. Also tabulated are 1.4~GHz radio fluxes from the selected galaxies and their IRAS fluxes.

An ordinary linear least squares fit is performed on this data to determine the FIR/RC ratio as defined by the data from Y01. The fit equation calculated in this paper is Equation~\ref{eqn:YunFit}:

%
\begin{equation}\label{eqn:YunFit}
\begin{array}{*{20}{c}}
{{\rm{log}}({{\rm{L}}_{{\rm{1}}.{\rm{4GHz}}}}){\rm{ = }}\left( {\frac{{0.995W}}{{Hz}}} \right)\cdot\log \left( {\frac{{{L_{60\mu m}}}}{{{L_ \odot }}}} \right) + \left( {\frac{{12.377W}}{{Hz}}} \right)}\\
{}
\end{array}
\end{equation}
%



For each $5''$~$\times$~$5''$ pixel of the SMC data sets, the value of the FIR/RC ratio is calculated. The sums of this ratio for each region is plotted on the graph with Y01's data set (Figure~\ref{fig:yuncorr}). The line described by the fit equation is extended through the SMC data sets presented in this paper to show how the fitted ratio of the Y01 data corresponds with this SMC data.  The correlation is evident in all five regions and for the whole SMC. The distance from Y01's regression fit line to the regional and complete SMC data  are very small: from $\Delta$=0.017~dex to $\Delta$=0.115~dex. These differences changed slightly depending on whether the background sources are included or not. The differences for each region and the whole SMC are shown in Table~\ref{tbl:yuncompare} including the change due to background sources.

For comparison, Figure~\ref{fig:yuncorr} also plots the LMC correlation calculated from data published on the LMC by \cite{hugh06}.  The distance to Y01's regression line for the LMC is calculated here to be $\Delta$=0.308~dex.

%

\begin{table}
\small
\caption{Comparison of the difference between the data fit to data from Yun \citep{yun01} and the current measurements. The contribution to the fit from the background sources is shown to contribute less than 10 percent to the fit discrepancy.}
\label{tbl:yuncompare}
\begin{tabular}{@{}cccc@{}}
\tableline  
Regions & Without BKG & With BKG & Difference \\
 & dex & dex & \%\ \\
\tableline
1&0.1159& 0.1153&-0.52\\
2&0.0431&0.0433&0.46\\
3&0.0465&0.0460&-1.08\\
4&0.0183&0.0172&-6.39\\
5&0.0297&0.0283&-4.94\\
SMC&0.0783&0.0783& 0.00\\
\tableline 
\end{tabular}
%
\end{table}

\subsection{Pixel by Pixel analysis}\label{s:PbyP}

Pearson's linear correlation coefficient calculation is performed between several pairs of data sets, Equation~\ref{eqn:Pearson}. For each pair of data sets, the data set with the smallest FWHM beamwidth is convolved with a Gaussian kernel to smooth it to the same FWHM as the lower resolution data. The sampling is then done at a spacing corresponding to the beamwidth of the lower resolution data to guarantee statistical independence of the samples. In this equation $f_N$ refers to the $N^{th}$ image, $f_{Ni}$ refers to the $i^{th}$ pixel of the $N^{th}$ image and $r_p$ is the Pearson's linear correlation coefficient.

\begin{equation}\label{eqn:Pearson}
{r_p} = \frac{{\sum {\left( {{f_{1i}} - \left\langle {{f_1}} \right\rangle } \right)\left( {{f_{2i}} - \left\langle {{f_2}} \right\rangle } \right)} }}{{\sqrt {\sum {{{\left( {{f_{1i}} - \left\langle {{f_1}} \right\rangle } \right)}^2}\sum {{{\left( {{f_{2i}} - \left\langle {{f_2}} \right\rangle } \right)}^2}} } } }}
\end{equation}

The error in this correlation is calculated as follows:

\begin{equation}\label{eqn:deltaPearson}
\Delta {r_p} = \frac{{\sqrt {1 - r_p^2} }}{{\sqrt {{n^2} - 2} }}
\end{equation}

The formal errors for these data sets are $\simeq$$10\% $ since $n$, the total number of points in a data set, is $ \sim$${10^3}$ when the resolution of the data sets are considered. Of course, the real error is possibly larger due to calibration errors and other systematic errors in the data.

Perfectly correlated images would have a coefficient of 1 and perfectly anti-correlated images would have a coefficient of -1. Table \ref{tbl:PearsonCorr} shows various pairs of data sets for the 5 regions and the correlations therein. Values above 0.7 or 0.8 in Table~\ref{tbl:PearsonCorr} indicates moderate to strong correlation. Those values are printed in bold face type.

Region~1 and Region~5 have the highest correlations with the FIR-\HI\ data sets and includes N\,19 and NGC\,267.  SMC SNR  position data from \cite{fili05} is used to determine the SNR population of each region.  Region~1 has the largest concentration of \HI\ for the SMC (see Table~ \ref{tbl:fluxes}) and a complex of 5 SNRs around SNR~B0045-734.  There are 4 other SNRs in Region~1. Region~5 contains no identified SNRs. Region~2 shows very little correlation in any of the data set pairs. There are two identified SNRs in Region~2. Region~3 is dominated by the \HII\ region N\,66 and NGC\,346, and shows good correlations in the FIR-RC data sets as well as the H$\alpha$ correlations with the RC and FIR data. Region~3 contains 12~SNRs. Region~4 shows moderate to strong correlations of the IR-CO data sets for the coolest dust. Region~4 is not well covered by the $70~\mu m$ or the CO data. There are no known SNRs in Region~4. It does, however, contain NGC\,456.

\begin{table}
\small
\caption{Pearson correlations for pairs of data sets. All results have a calculated
error of $\simeq10\%$. For each pair of data sets, the data set with the smallest FWHM beamwidth is convolved with a Gaussian kernel to smooth it to the same FWHM as the lower resolution data. The sampling is then done at a spacing corresponding to the beamwidth of the lower resolution data. Values above $0.7$ are emphasized as having good correlation.}

\label{tbl:PearsonCorr}
\begin{tabular}{@{}lccccc@{}}
\tableline  
 & & & Regions & & \\
\tableline
Data Sets & 1 & 2 & 3 & 4 & 5\\
\tableline
RC 21 cm & & & & & \\
$60~\mu m$&	0.620&	0.686&	\bf{0.825}&	0.444&	0.263\\
$70~\mu m$&	0.573&	0.513&	0.585&	0.484&	0.193\\
$100~\mu m$&	0.574&	0.607&	\bf{0.752}&	0.410&	0.201\\
$160~\mu m$&	0.587&	0.458&	0.656&	0.515&	0.168\\
\tableline
RC 13~cm & & & & & \\
$60~\mu m$&	0.609&	\bf{0.700}&	\bf{0.815}&	0.501&	0.304\\
$70~\mu m$&	0.610&	0.628&	\bf{0.752}&	\bf{0.720}&	0.127\\
$100~\mu m$&	0.569&	0.638&	\bf{0.740}&	0.461&	0.244\\
$160~\mu m$&	0.629&	0.577&	0.661&	0.616&	0.129\\
\tableline						
CO & & & & & \\
RC 21~cm&	0.272&	0.059&	0.311&	0.422&	-0.255\\
RC 13~cm&	0.296&	0.073&	0.305&	0.347&	0.032\\
$60~\mu m$&	0.531&	0.130&	0.318&	\bf{0.759}&	0.116\\
$70~\mu m$&	0.531&	0.142&	0.405&	0.471&	0.190\\
$100~\mu m$&	0.554&	0.170&	0.378&	\bf{0.769}&	0.146\\
$160~\mu m$&	0.599&	0.168&	0.479&	\bf{0.896}&	0.181\\
\HI&	0.304&  0.205&	0.173&	0.444&	0.293\\
$H\alpha$&	0.312&	0.018&	0.307&	0.291&	-0.055\\
\tableline	
\HI & & & & & \\	
RC 21~cm&	0.404&	0.255	&0.059&	0.092	&0.137\\					
RC 13~cm&	0.443&	0.307	&0.044&	0.137	&0.109\\
$60~\mu m$&	\bf{0.762}&	0.584&	0.197&	0.394&	\bf{0.748}\\
$70~\mu m$	&\bf{0.709}&	0.545&	0.209&	0.226	&\bf{0.790}\\
$100~\mu m$&	\bf{0.810}&	0.680&	0.338&	0.472&	\bf{0.833}\\
$160~\mu m$&	\bf{0.773}&	\bf{0.708}&	0.401&	0.506&	\bf{0.857}\\
\tableline	
$H\alpha$ & & & & & \\	
RC 21~cm&	0.377&	0.659&	\bf{0.859}&	0.563	&0.329\\						
RC 13~cm&	0.273&	0.597&	\bf{0.830}&	\bf{0.722}	&0.237\\
$60~\mu m$&	0.205&	0.327&	\bf{0.816}&	0.516&	0.408\\
$70~\mu m$	&0.387	&0.290&	0.638	&0.535	&0.018\\
$100~\mu m$&	0.158&	0.297&	\bf{0.731}&	0.475&	0.350\\
$160~\mu m$&	0.438&	0.247&	0.661&	0.179&	0.020\\
\tableline 
\end{tabular}
%
\end{table}

%
\begin{figure}
\includegraphics[width=\columnwidth]{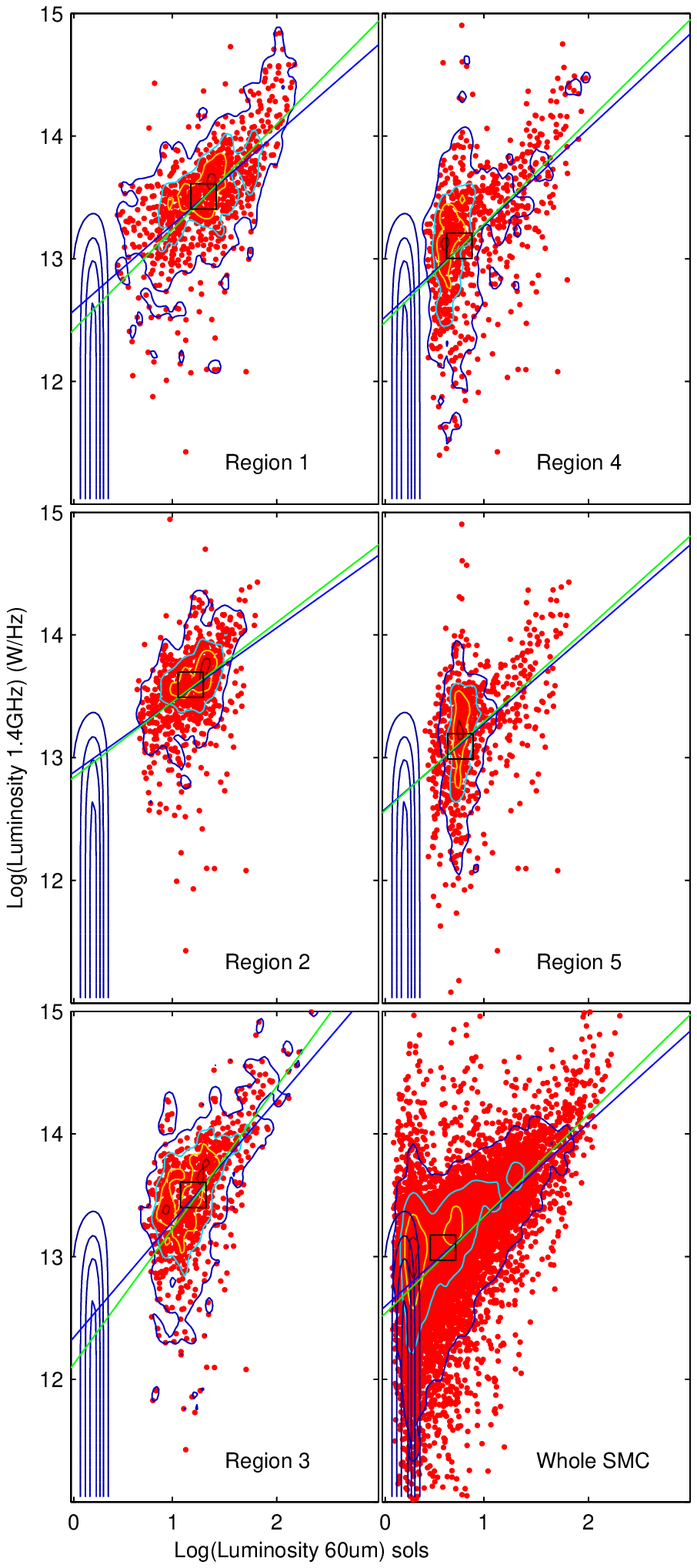}
\caption{Scatter plot of the regions and the whole SMC. For each pair of data sets, the data set with the smallest FWHM beamwidth is convolved with a Gaussian kernel to smooth it to the same FWHM as the lower resolution data. The sampling is then done at a spacing corresponding to the beamwidth of the lower resolution data. The two lines in each plot are the ordinary least squares and weighted least squares fits to the data. The OLS is in green, the WLS in blue. The contours in the data plot show the density of data points and the Gaussian distribution of the background noise. Those contours are at 5, 30, 60, and 90\% of the peak data density. The position of the square is the average of the data set.}
\label{fig:scatterSMC}
\end{figure}

Figure~\ref{fig:scatterSMC} shows the scatter plots of the 21~cm continuum and $60~\mu m$ IR data for the pixels of all five regions and the whole SMC. For each pair of data sets, the data set with the smallest FWHM beamwidth is convolved with a Gaussian kernel to smooth it to the same FWHM as the lower resolution data. The sampling is then done at a spacing corresponding to the beamwidth of the lower resolution data to guarantee statistical independence. It also shows two different Least Squares Fits to each data set. The blue line is the Ordinary Least Squares Fit, OLS and the green line is the Weighted Least Squares Fit (WLS) calculated using the Matlab function $lscov$  \citep{matl10}. The OLS assumes the error of each data point is the same and consequently weights every data point equally. The WLS uses a weighting for each data point which here is calculated as the square root of the sum of the squares of the values of each axis linearly weighted from 0.01 to 1 scaled to the value of the data point from the minimum value to the maximum value of each data set.

The difference in the slope and offset of these fit lines and the averages are shown in Table~\ref{tbl:AvgOlsWls}. The last row contains the differences in percent of the OLS slope and the WLS slope of the fitted lines. It is expected that the OLS fit would be a better representation of the entire data set whereas the WLS fit is weighted to lower the contribution of data with larger calculated errors and would describe the strongest data without the effect of outlier points.

Strong to moderate Pearson correlations in FIR-RC, FIR-\HI, and IR-H$\alpha$ in Regions~1 and 3 seen in Table~\ref{tbl:PearsonCorr} appear to show the effect of two populations of data in those regions. The range of OLS slopes in the regions is 0.594 to 0.978 while the WLS slopes range from 0.638 to 1.144 over the regions. In every region the OLS slope is flatter that the WLS slope. The set of contour lines to the left of the figures is from the background noise using measurements from the north east corner of the whole SMC data set which are not considered to be part of the SMC. The measured average value and the RMS noise are used to calculate a background Gaussian noise distribution and the results of this distribution are used to create the noise contours for the background. Also shown are four contours which are created by binning the luminosity data into a 100$\times$100 bin array. The contours shown are at 5, 30, 60 and 90\% of the maximum bin count. Visually examining these contours indicates that 90\% of the data is found with values quite symmetric to the fitted lines, both OLS and WLS. This suggests that the WLS is a better representative of the data in the regions. The Gaussian noise distribution calculated for the measured background noise uses the same scaling. It is clear that there is not a significant contribution by noise to the scatter plot for any of the five regions. However, the whole SMC plot includes portions of sky which are not part of the SMC and contain only noise so that there is considerable overlap seen in the lower luminosity values of the scatter plot with the noise contours. The plotted square in each data set is the location of the average pixel value.

\begin{table}
\small
\caption{For each pair of data sets, the data set with the smallest FWHM beamwidth is convolved with a Gaussian kernel to smooth it to the same FWHM as the lower resolution data. The sampling is then done at a spacing corresponding to the beamwidth of the lower resolution data. Data set averages for RC 21~cm, Log(Luminosity 1.4~GHz)(W/Hz), and FIR, Log(luminosity $60~\mu m$) sols, and fit parameters to y=ax + b using Ordinary Least Squares fit, OLS, and Weighted Least Squares fit, WLS. These numerical values are shown graphically in Figure~\ref{fig:scatterSMC}. The differences in slope values are also shown with the WLS always larger than the OLS value. For comparison, Y01's data as fitted here using OLS gives $a = 0.995$ and $b = 12.38$.}
\label{tbl:AvgOlsWls}
\begin{tabular}{@{}lcccccc@{}}
\tableline  
 & & & Regions & & \\
\tableline
 & 1 & 2 & 3 & 4 & 5 & SMC\\
\tableline
RC&	13.51&	13.60&	13.50&	13.11&	13.10& 13.07\\
FIR&	1.296&	1.169&	1.195&	0.753&	0.762& 0.590\\
OLS a&	0.730&	0.594&	0.978&	0.776&	0.725& 0.755\\
WLS a&	0.846&	0.638&	1.144&	0.829&	0.752& 0.819\\
OLS b&	12.55&	12.86&	12.31&	12.50&	12.56& 12.57\\
WLS b&	12.40&	12.82&	12.09&	12.46&	12.55& 12.51\\
$\Delta a \%$& 13.7 & 6.8 & 14.5 & 6.4 & 3.6 & 7.8 \\
\tableline 
\end{tabular}
%
\end{table}

\subsection{SMC FIR/RC ratio map}
 \label{s:qplot}
%
\begin{figure}
\includegraphics[width=\columnwidth]{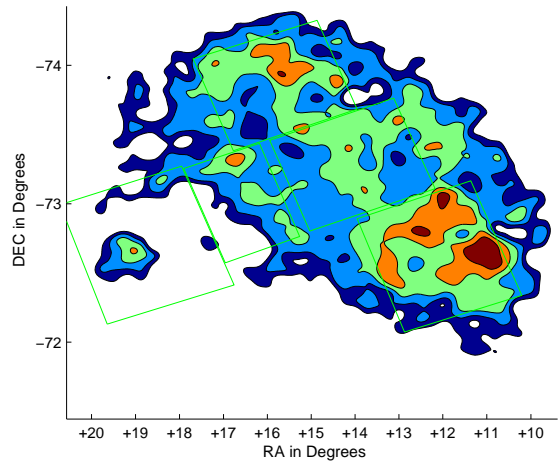}
\caption{This figure shows the q value plot of the SMC which has been convolved with a Gaussian kernel of 108 pc FWHM.
Red  $>2.95>$ yellow $>2.75>$ green $>2.55>$ light blue  $ >2.35 >$ dark blue  $ >2.15 > $  white.}
\label{fig:qSmeared}
\end{figure}

In order to relate particular physical structures in the SMC with variations in the FIR/RC ratio, Figure~\ref{fig:qSmeared} is calculated showing the log ratio of FIR and 21~cm RC data. The q value map is calculated using the formula from \cite{helo88} with the IRAS $60~\mu m$ and $100~\mu m$ data sets. The FIR value is calculated in equation~\ref{eqn:FIR}.

\begin{equation}\label{eqn:FIR}
{\left( {\frac{{FIR}}{{W{m^{ - 2}}}}} \right) = 1.26 \cdot {10^{ - 14}}\left( {\frac{{2.58{S_{60\mu m}} + {S_{100\mu m}}}}{{Jy}}} \right)}
\end{equation}

The q values of the pixels are calculated with Equation \ref{eqn:q}.

\begin{equation}\label{eqn:q}
q = \log \left( {\frac{{FIR}}{{3.75 \cdot {{10}^{12}}Wm{}^{ - 2}}}} \right) - \log \left( {\frac{{{S_{RC}}}}{{W{m^{ - 2}}H{z^{ - 1}}}}} \right)\
\end{equation}

 The average value of the ratio is q=2.65. This suggests that the SMC has a slight FIR excess when compared to Y01's average figure of q=2.34. Approximately 98\% of Y01's galaxies are within a factor of 5 of 2.65, $\sim$1.6 to $\sim$3.0. Figure~\ref{fig:qSmeared} shows the distribution of q values across the SMC from q~$<$2.15 to q~$>$2.95.

 Region~1, with an average q of 2.80, has an a large portion of its area with q $\geq$2.65. This suggests that there is abundant dust in the area. Relatively high Pearson correlations of FIR with \HI\ are also seen in Table~\ref{tbl:PearsonCorr} for Region~1. There are 9 SNRs in this region. The strongest RC feature in this region is N\,19 containing SNR~B0045-734 \citep{payn04} \citep{craw11,wong11,wong11a,wong12}. The other regions have extended areas where the RC excess is present.

 Region~2, with an average q of 2.58, has a large portion of its area with q $\leq$2.65. With two SNRs in this region, it could be inferred that there is not much star-forming activity in this region so that the dust is cooler.

 Region~3, with an average q of 2.70, has 12 SNRs with similar areas with q $<$2.65 and q $>$2.65. The high star formation rate leads to the very high Pearson correlation seen in all of the IR data maps with the 21~cm data. $H\alpha$ correlates also very well with the RC data and the FIR data.

 Region~4, with an average q of 2.75, is dominated by background radio source SMC~B0109-7330. It has a portion of its area with significant dust and gas in the eastern part of the region. It is the only region that has strong correlation of CO with the IR emission in the Pearson correlation results.

 Region~5, with q of 2.73, has very small total energy emission, 0.2 -- 0.3 of the average of the other 5 regions. Scaled for area, it is $\sim$0.4 - $\sim$0.6 of the average.

\subsection{Wavelet analysis}
 \label{s:wavelet}

Wavelets can be used as a tool for scaling analysis \citep{fric01}. It is a technique that convolves the data with a family of self-similar analyzing functions that depend on scale and location. The family of analyzing functions is defined by dilations and translations of the analyzing function, also called the mother function. In two dimensions, the continuous wavelet transformation can be written as follows:

\begin{equation}\label{eqn:wtrans}
W(a,\vec x) = \frac{1}{{{a^\kappa }}}\int_{ - \infty }^\infty  {\int_{ - \infty }^\infty  {f(\vec x'){\Psi ^*}\left( {\frac{{\vec x' - \vec x}}{a}} \right)d\vec x'} }
\end{equation}

In this expression, $\vec x = \left( {x,y} \right)$ and $f\left( {\vec x} \right)$ is the two dimensional data set for which the Fourier transform exists, ${\Psi ^*}\left( {\vec x} \right)$ is the complex conjugate of the analyzing function, $a$ is the scale factor, and $\kappa $ is the normalizing parameter.

The analyzing function used here is the Pet Hat (PH) function \citep{fric01}. This function is defined in terms of its Fourier transform as follows:

\begin{equation}\label{eqn:ph}
\begin{array}{c}
\hat \Psi \left( {\vec k} \right) = {\cos ^2}\left( {\frac{\pi }{2}{{\log }_2}\frac{k}{{2\pi }}} \right){\rm{ }}:{\rm{ }}\pi  < \left| {\vec k} \right| < 4\pi \\
{\rm{ =   0                          : otherwise}}
\end{array}
\end{equation}

This function specifies an annulus in Fourier space with a median radius of $2\pi$ and is non-zero only for the specified annulus.

\subsection{Wavelet cross correlation}
 \label{s:wavecross}

Given a two dimensional data set, the energy of the transformation as a function of scale factor $a$ over the entire plane is:

\begin{equation}\label{eqn:wtrans}
M\left( a \right) = \int_{ - \infty }^\infty  {\int_{ - \infty }^\infty  {{{\left| {W(a,\vec x)} \right|}^2}} } d\vec x
\end{equation}

The wavelet cross correlation coefficient as a function of $a$ is calculated as follows:

\begin{equation}\label{eqn:wcross}
{r_w}\left( a \right) = \frac{{\iint {{W_1}\left( {a,\vec x} \right)W_2^*\left( {a,\vec x} \right)d\vec x}}}{{\sqrt {\left[ {{M_1}\left( a \right){M_2}\left( a \right)} \right]} }}
\end{equation}

where the subscripts refer to the source data sets used in the cross correlation.

The error in the cross correlation is calculated as follows:

\begin{equation}\label{eqn:rwerr}
\Delta {r_w}\left( a \right) = \frac{{\sqrt {1 - r_w^2(a)} }}{{\sqrt {{{\left( {\frac{L}{a}} \right)}^2} - 2} }}
\end{equation}

where $L$ is the linear size of the data set. For the data used here $\Delta {r_w}\left( a \right)$ is $ \lesssim 15\% $.

For each pair of data sets used in wavelet cross correlation, the data set with the smallest FWHM beamwidth is convolved with a Gaussian kernel to smooth it to the same FWHM as the lower resolution data. The sampling is then done at a spacing corresponding to the beamwidth of the lower resolution data. Figure~\ref{fig:wave15} shows wavelet correlations of several data sets with the 21~cm RC in all 5 regions. The smallest scale that can be used to analyze the data is the resolution of the data. The largest scale is dependent on the overall size of the data set. The smallest linear dimension of the data set should be no less than four times the largest scale factor used in the analysis. The resolution of the 21~cm data is 90$''$ (27~pc) and the finest resolution calculated is 80$''$ (24~pc). The correlation is performed with IRAS $60~\mu m$ and $100~\mu m$ data with 120$''$ (36~pc) resolution, Spitzer $70~\mu m$ and $160~\mu m$ data with 18$''$ (5.4~pc) and 40$''$ (12~pc) resolution respectively, \HI\ data with 98$''$ (30~pc) resolution, $H\alpha$ with 3$''$ (1~pc) resolution, and CO data with 156$''$ (46~pc) resolution. For the data sets used here, the maximum valid scale value for the wavelet cross correlation is 794$''$ (238~pc) for Regions 1-4 and 501$''$ (150~pc) for Region~5 except for the CO data.

The CO map is much more irregular than the other data sets. Region~1 is close to being fully covered with the cross correlation results beginning at 125$''$. Region~2 is not well covered and has a large hole in the coverage in the center of that region. The only scale factors that satisfy validity requirements are at 125$''$ and 200$''$, both of which are shown. Region~3 is also close to being fully covered with the cross correlation results beginning at 125$''$. Region~4 is split into two smaller areas. Interpreting the wavelet cross correlation in the region is problematic. Region~5 is just a sliver of data that is too small and irregular to use in a wavelet correlation. Wavelet correlations of Regions~4 and 5 are not calculated.

\subsection{Wavelet cross correlation for FIR with RC}
 \label{s:firrccross}

%
\begin{figure}
\includegraphics[width=\columnwidth]{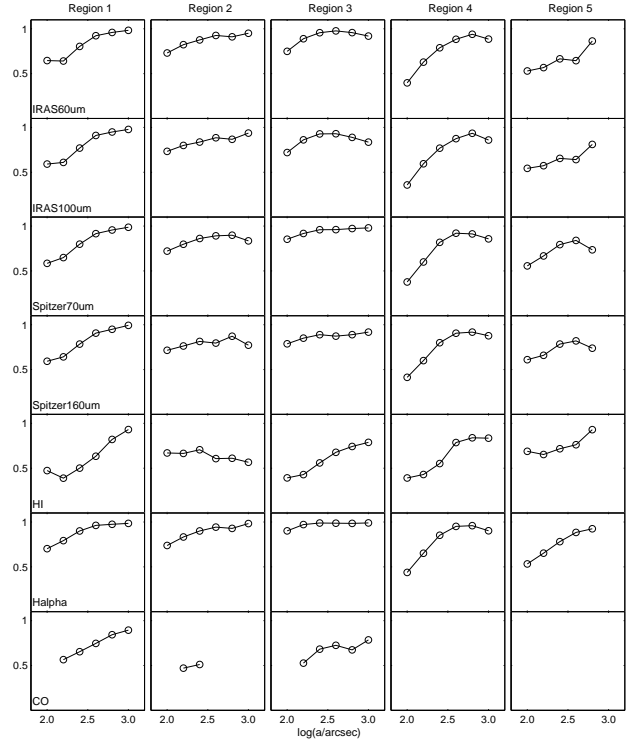}
\caption{21~cm RC wavelet cross correlation plots for Region 1 through 5 with data from several other data sets. The data points for Regions 1-4 are calculated at scale factors of 80$''$, 125$''$, 200$''$, 316$''$, 501$''$, and 794$''$ which correspond to 24~pc, 37~pc, 60~pc, 95~pc, 150~pc and 238~pc respectively. The data set from Region~5 is too small to support wavelet cross correlations of 794$''$ so correlations are only shown up to 501$''$. Coverage over the map of CO data is not complete except for Regions~1 and 3.}
\label{fig:wave15}
\end{figure}

\begin{table}
\small
\caption{21~cm RC wavelet cross correlation scale factors in pc for Region~1 through 5 with data from several other data sets. Listed are the minimum scale factors in pc for which the wavelet cross correlations are considered excellent, which  is taken to be $\geq0.75$. Note that the $\bf{Bold Face}$ values are limited by the beamwidth of one of the data sets.}
\label{tbl:WavCrossValues}
\begin{tabular}{@{}lccccc@{}}
\tableline  
 & & Regions & & \\
\tableline
 & 1 & 2 & 3 & 4 & 5 \\
\tableline
IRAS $60~\mu m$&	53.2& \bf{35.3}& \bf{35.3}&	51.5&	129.8\\
IRAS $100~\mu m$&	57.3& \bf{35.3}&\bf{35.3}&	55.9&	136.1\\
Spitzer $70~\mu m$&	52.4& \bf{28.3}& \bf{24.0}&	50.6&	50.3\\
Spitzer $160~\mu m$&	54.5&	33.4& \bf{24.0}&	52.6&	53.2\\
\HI&	129.8& --- &	170.1&	88.7&	83.6\\
H$\alpha$&	30.6& \bf{25.3}& \bf{24.0}&	45.9&	53.2\\
CO& 97.0 & --- & 229.0 & --- & --- \\
\tableline 
\end{tabular}
%
\end{table}

Table \ref{tbl:WavCrossValues} shows the smallest scale factors for which excellent wavelet cross correlation is measured. \cite{fric01} regarded 0.75 or above as representing an excellent correlation between images and will be referred to here as the correlation threshold. The scale factors are extracted from the data shown in Figure~\ref{fig:wave15}. The data clearly shows strong morphological correlations of dust emission in the FIR with the 21~cm continuous emission from the defined regions of the SMC down to beamwidth limited minimum scale factor, $\sim35~pc$. In Regions~2 and 3, the limit of the correlations is primarily due to the data resolution. Table~\ref{tbl:WavCrossHiRes} shows the wavelet correlations extracted from Figure~\ref{fig:wave15h} which correlates the Spitzer IR data with the higher resolution 3~cm and 6~cm data, \citep{dick10}. The wavelet cross correlation with the Spitzer data shows that the FIR/RC wavelet correlation scale factors extend down to $\sim$15~pc. Within each region, all of the FIR bands show similar correlation patterns with RC data with respect to scale factor. The coverage of the regions is quite good for the Spitzer 160~$\mu m$ data but less so for the Spitzer 70~$\mu m$ data particularly in Regions 1, 3, and 4. There are also some unexpected differences between the 3~cm and 6~cm correlations with the FIR data.

\subsection{Wavelet cross correlation for other bands}
 \label{s:otherccross}

H\,I cross correlation is seen in Figure~\ref{fig:wave15} with 21~cm RC in Region~1. The correlation initially dips before beginning a slow rise to the correlation threshold at $\sim$130~pc. Region~2 never achieves correlation threshold and Region~3 doesn't until $\sim$170~pc. Regions~4 and 5 have approximately the same scale factor of $\sim$85~pc.

\begin{figure}
\includegraphics[width=\columnwidth]{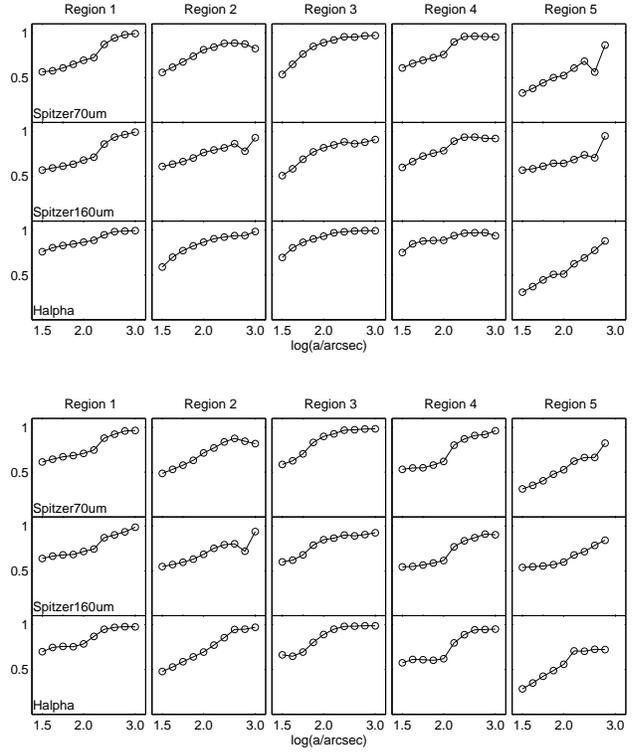}
\caption{6~cm (top) and 3~cm (bottom) wavelet cross correlation plots for Regions 1 through 5 with the Spitzer FIR data sets and $H\alpha$ data. These graphs extend the cross correlation calculations down to 9pc. The data points are calculated at scale factors of 31$''$, 40$''$, 50$''$, 63$''$, 80$''$, 125$''$, 200$''$, 316$''$, 501$''$, and 794$''$ which correspond to 9~pc, 12~pc, 15~pc, 19~pc, 24~pc, 37~pc, 60~pc, 95~pc, 150~pc and 238~pc respectively. Region 5 scale factors are limited to a maximum of 501$''$, 150~pc due to the small size of the region.}
\label{fig:wave15h}
\end{figure}

H$\alpha$ correlations from Figure~\ref{fig:wave15} in Table~\ref{tbl:WavCrossValues} with 21~cm RC are essentially at the resolution of the \HI\ data at $\sim$25~pc for Regions 1 through 3. Regions 4 and 5 have correlation threshold at larger scale factors of $\sim$85~pc. High-resolution H$\alpha$ correlations from Figure~\ref{fig:wave15h} in Table \ref{tbl:WavCrossHiRes} with 3~cm and 6~cm shows that the H$\alpha$ wavelet cross correlation extends down to much smaller scale factors from 9~pc to 33~pc for Regions~1 through 4.

The CO correlations are very sparse since the map of the data is rather incomplete. Region~1 had correlation threshold at 97~pc and above.  No significant correlation is detected in Region~2. Region~3 had correlation threshold at 229~pc. No other correlations are possible.

%

\begin{table}
\small
\caption{6~cm and 3~cm RC high resolution wavelet cross correlation scale factors in pc for region~1 through 5 with data from Spitzer and $H\alpha$ data sets. Listed are the minimum scale factors in pc for which the wavelet cross correlations are considered excellent, which is taken to be $\geq0.75$.}
\label{tbl:WavCrossHiRes}
\begin{tabular}{@{}lccccc@{}}
\tableline  
 & & 6~cm RC& & \\
\tableline
 Region& 1 & 2 & 3 & 4 & 5 \\
\tableline
Spitzer 70~$\mu m$&	43.2&	19.8&	14.7&	23.1&	137.3\\
Spitzer 160~$\mu m$&	45.2&	22.7&	17.7&	17.6&	114.1\\
$H\alpha$&	9.0&	14.0&	10.4&	9.1&	87.5\\
\tableline  
 & & 3 cm RC& & \\
\tableline
Spitzer 70~$\mu m$&	37.9&	27.6&	16.4&	33.3&	129.4\\
Spitzer 160~$\mu m$&	39.7&	36.8&	17.6&	35.5&	80.6\\
$H\alpha$&	12.6&	32.1&	17.10&	33.8&	----\\
\tableline 
\end{tabular}
%
\end{table}

\section{Discussion}
 \label{s:Discussion}

The FIR/RC correlation coefficient, q, determined in this study is very close to the expected value for normal galaxies even though the SMC is structurally different from the spiral galaxies which constitute most of the population on which the correlation is based. Spirals generally have a thick underlying non-thermal disk. The expected thermal fraction for the disk of a spiral galaxy is $ \sim$0.1 \citep{cond92}. The three regions that cover the SMC bar measured here all have high thermal ratios of $\sim$0.54 to $\sim$0.95 (see Figure~\ref{fig:fth}). The existence of the bar, however, is questioned by \cite{zari00}. Their argument centers around the hydrodynamic interaction between gaseous components providing the physical mechanism for the SMC morphology. They further suggest that the SMC is actually spheroidal with highly irregular recent star formation. The overall thermal fraction of the SMC is measured here using 1.4~GHz and 4.8~GHz data to have a value of $\sim$0.43 which is higher than previously published values of 0.15-0.40 for the SMC by \cite{lois87} using 1.4 GHz and 2.3~GHz data but lower than the value of 0.71 measured by \cite{isra10} using 5~GHz and 10~GHz data. However, each of the SMC regions measured here except Region 2 has a thermal fraction of $>$0.71.

The triggers for star formation events in massive spirals are believed to be density waves \citep{beck00}. In the SMC, there is not enough mass to support such a mechanism so that the trigger in the SMC is likely to be stochastic which leads to a discontinuous formation history with bursts of star formation lasting short periods of time \citep{klein86}. With only two SNRs detected in Region~2, there is clearly no evidence of a galactic nucleus with its traditional high star formation rate. Figure~\ref{fig:qSmeared} shows the distribution of q values in the SMC and there is no indication of the q value decreasing from a $''$nucleus$''$ to the edge of the galaxy as \cite{helo93} expect in their $''$leaky box$''$ FIR/RC model. Figure~\ref{fig:qSmeared} does show that most of the SMC bar is consistent with q values within Y01's expected limits. The depth of the SMC extends 5-10~kpc along the line of sight, \citep{mart89}. The measured thermal fraction leads to the conclusion that there are not enough ${\rm{CR}}{{\rm{e}}^ - }$s to generate higher non-thermal emissions and/or that the magnetic field strength is too low in most regions. The published values of the magnetic field are $5-10~\mu $G along the bar \citep{ye91}. \cite{sree91} concluded that the cosmic ray energy density level in the SMC is three to five times too low to be in equilibrium with the cosmic ray sources. This low cosmic ray density is described by \cite{klein84} to be a result of the internal pressure from the ${\rm{CR}}{{\rm{e}}^ - }$s.  This causes rapid diffusion of the ${\rm{CR}}{{\rm{e}}^ - }$s out of the star-forming regions. This is consistent with the observation that the bar has primarily thermal RC emission in Region~1 and Region~3 where there is high density \HI\ and star formation activity. Region~2 may be populated primarily with ${\rm{CR}}{{\rm{e}}^ - }$s which have diffused out of the two star-forming regions in Region~1 and Region~3.

\subsection{FIR/RC Correlation Models}\label{s:Models}

The FIR/RC correlation models examined here start with the assumption that the source of relativistic electrons and ionizing photons is the supernova rate. UV photons from massive stars are absorbed by dust which re-emit the energy in the FIR. The relativistic electrons from the SNRs interact with magnetic fields or interact with ionized gas to produce RC radiation.

\cite{voel89} proposed a calorimeter theory which assumes that all of the ionizing photons are absorbed by dust and all of the energy of the ${\rm{CR}}{{\rm{e}}^ - }$s are dissipated by RC emission within the galaxy. This does not appear to be consistent with the SMC since \cite{sree91} determined that most of the ${\rm{CR}}{{\rm{e}}^ - }$s escape the SMC.

Applying the $''$leaky box$''$ model of \cite{helo93} most of the ionizing photons and ${\rm{CR}}{{\rm{e}}^ - }$s would escape the SMC. The observational prediction of the model is that q decreases radially from the center of the galaxy to the edges of the galaxy. Regions~1 and 3 are the main star-forming regions in the SMC at the present time. Star formation in Region~2 must be suppressed since the $H_2$ content is the lowest of any region except for Region 5, see Table~\ref{tbl:fluxes}. Due to the irregular nature of the SMC, different position angles from these star-forming areas point towards very different environments as seen in Figure~\ref{fig:qSmeared}.   This does not appear to match the gradual decrease in q modeled by Helou and Bicay particularly since neither a nucleus nor a disk can be identified in the SMC.

\cite{hoer98} (referred to hereafter as H01) published a model which decomposes both the RC emission and FIR emission into two parts. M\,31 is used as their test case. H01 selection of M\,31 relied on four criteria:

\begin{enumerate}
\item M\,31 is close to the Galaxy and well resolved in both radio and in the FIR.
\item The radio emission is dominated by the non-thermal component.
\item The FIR is dominated by cool dust emission.
\item M\,31 has a low star formation rate of about an order of magnitude below the Galaxy.
\end{enumerate}

Of these criteria, the SMC is also close enough for well resolved FIR and RC measurements. But the SMC is not dominated by the non-thermal RC component with the overall SMC non-thermal emission at 57\% and a minimum non-thermal emission of only 5\% in Region~3, see Figure~\ref{fig:fth}.

The supernova rate (Rs) for the SMC is about 1/350  years, ${\rm{R}}{{\rm{s}}_{{\rm{SMC}}}} = 2.8 \cdot {10^{ - 3}}SN \cdot y{r^{ - 1}}$  \citep{fili98} whereas the Galaxy has a supernova rate of about 1/50 years,  ${\rm{R}}{{\rm{s}}_{{\rm{Galaxy}}}} = 20 \cdot {10^{ - 3}}SN \cdot y{r^{ - 1}}$ \citep{dieh06}. Dividing these rates by the galaxies' respective volumes shows that  the  supernova rate per unit volume is higher in the SMC than in the Galaxy:

 \begin{equation}\label{eqn:snr-smc}
\frac{{{\rm{R}}{{\rm{s}}_{{\rm{SMC}}}}}}{{{{\rm{V}}_{{\rm{SMC}}}}}} \approx \frac{{2.8 \cdot {{10}^{ - 3}}SN \cdot y{r^{ - 1}}}}{{15kp{c^3}}} \approx 0.19\frac{{SN}}{{kp{c^3}kyr}}
\end{equation}

 \begin{equation}\label{eqn:snr-galaxy}
\frac{{{\rm{R}}{{\rm{s}}_{{\rm{Galaxy}}}}}}{{{{\rm{V}}_{{\rm{Galaxy}}}}}} \approx \frac{{20 \cdot {{10}^{ - 3}}SN \cdot y{r^{ - 1}}}}{{1200kp{c^3}}} \approx 0.017\frac{{SN}}{{kp{c^3}kyr}}
\end{equation}

Whereas M\,31 has a star formation rate of an order of magnitude below the Galaxy with those two galaxies approximately the same volume, the SMC has a volumetric supernova rate an order of magnitude above the Galaxy or two orders of magnitude above M\,31.

The SMC does not appear to be consistent with the criteria needed to apply H01's decomposition model.

\subsection{Decomposition of the SMC RC into thermal and non-thermal emission}
 \label{s:fth}

Data from the Parkes single dish telescope are used for the RC thermal fraction calculations \citep{fili97} due to its superior overall flux-density accuracy. The Parkes data were converted to a data map with 5$''$ resolution using the Miriad regrid program to match the 6~cm data set. The thermal fraction of the emissions at 21~cm are calculated using equation \ref{eqn:fth} from \cite{nikl97}. $S_{6cm}$ and $S_{21cm}$ are the measured flux densities of the regions.

\begin{equation}\label{eqn:fth}
\frac{{{S_{6cm}}}}{{{S_{21cm}}}} = {f_{th}}{\left( {\frac{{{\nu _{6cm}}}}{{{\nu _{21cm}}}}} \right)^{ - {\alpha _{th}}}} + {f_{nth}}{\left( {\frac{{{\nu _{6cm}}}}{{{\nu _{21cm}}}}} \right)^{ - {\alpha _{nth}}}}
\end{equation}

\begin{figure}
\includegraphics[width=\columnwidth]{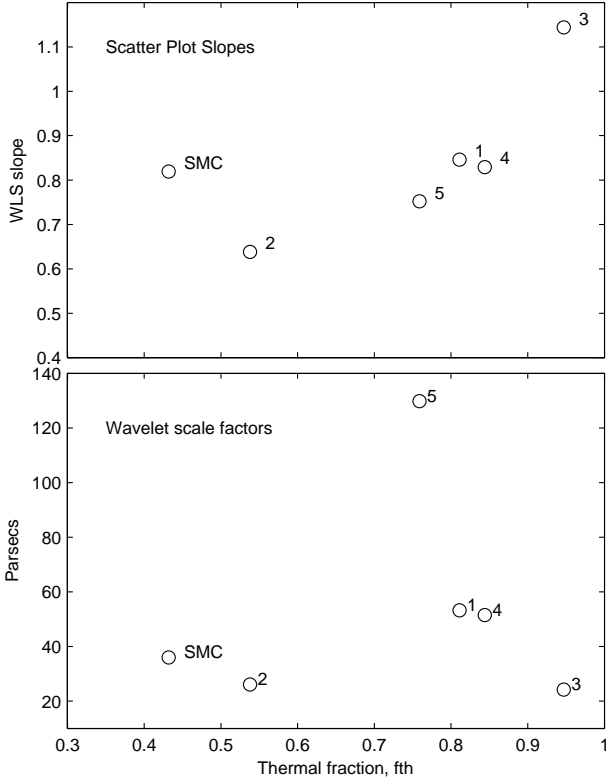}
\caption{Top Panel: WLS slope of 21~cm RC verses $60~\mu m$ data plotted against the thermal fraction. The thermal fractions plotted are 0.81, 0.54, 0.95, 0.84, and 0.76 for Regions~1 through 5. The SMC estimate is 0.43. Second Panel: Minimum wavelet cross correlation scale factor for 21~cm RC with $60~\mu m$ that had excellent correlation For Regions~1 through 5, those scale factors are 53.2~pc, 28.3~pc, 24.0~pc 51.5~pc, and 129.8~pc. The Region~2 and 3 are from the higher resolution Spitzer $70~\mu m$ data. For the ${3^ \circ } \times {3^ \circ }$ central portion of the SMC, excellent wavelet correlation is seen down to a scale factor of 36~pc.}
\label{fig:fth}
\end{figure}

In equation \ref{eqn:fth}, $f_{th}$ is the thermal fraction, $f_{nth}$ is the non-thermal fraction, ${f_{nth}} = (1 - {f_{th}})$. The thermal spectral index used is $\alpha _{th}=0.1$. The non-thermal spectral index is assigned a value of $\alpha _{nth}=1.09$ after \cite{bot10}. The thermal fractions calculated are 0.81, 0.54, 0.95, 0.84, and 0.76 for Regions~1 through 5 and 0.43 for the SMC, see Figure~\ref{fig:fth}. The thermal fraction calculation assumes that the synchrotron emission spectrum can be described by a single power law and that the ISM is optically thin. This assumption can introduce a bias into the calculated thermal fraction if the spectral index is not constant. The index is expected to steepen in locations that are increasingly remote from the area where the ${\rm{CR}}{{\rm{e}}^ - }$s were initially accelerated since the energy loss of ${\rm{CR}}{{\rm{e}}^ - }$s is $ \sim {E^2}$. Conversely, in star-forming regions the thermal fraction will likely be overestimated as the spectral index is flatter than in other regions.  However, the entire SMC as well as each region shows approximately 4 to 9 times the expected thermal fraction of 0.1 \citep{cond92} which suggests that ${\rm{CR}}{{\rm{e}}^ - }$s are not necessarily the source of the RC energy that makes the FIR/RC correlation so consistent.

%

\subsection{Regional Decomposition of the SMC }
 \label{s:fth-reg}

The top panel of Figure~\ref{fig:fth} shows the relationship of the slope of the Weighted Least Squares Fit (WLS) of 21~cm RC verses 60~$\mu m$ emissions from Table \ref{tbl:AvgOlsWls} plotted against the estimated thermal fractions. The regions show a strong tendency towards increasing WLS slope with increasing thermal fraction.

Region~3 has the highest WLS slope, the highest thermal fraction, and the highest 21~cm flux density of the five regions. It also has the best Pearson correlation, Table \ref{tbl:PearsonCorr}, of any region for RC 21~cm with all of the IR data sets, excellent Pearson correlation of 13~cm RC with the higher temperature IR emissions of $60~\mu m{\text{ and }}70~\mu m$ and excellent Pearson correlation of H$\alpha$ with RC 21~cm, 13~cm, IR $60~\mu m, 70~\mu m$ and $160~\mu m$ covering cold dust to warm dust emitters. The locations of the high \HI\ concentration correlates with the IR emissions by way of the high Pearson correlation indicating that the \HI\ and dust are co-located. It has excellent wavelet cross correlation of the 21~cm with $60~\mu m$ down to the resolution limit of 35~pc. Figure~\ref{fig:wave15h} shows the wavelet cross correlation of the higher resolution FIR and RC data ($70~\mu m$ and $160~\mu m$ Spitzer data with 6~cm and 3~cm RC data), indicating that the correlation extends down to 15pc. This region includes N\,66 and NGC\,346 which is an \HII\ region with embedded stellar association \citep{bica00} and the strongest star-forming region in the SMC \citep{cign11} with 12 SNRs identified.

The next two regions in terms of WLS slope and thermal fraction are Regions~1 and 4. They have WLS slope values and thermal fractions very similar to each other with WLS slope values, $ \thicksim 0.85$, and thermal fractions, $ \thicksim 0.82$. Region~1 contains N\,19 and NGC\,267, and Region~4 contains NGC\,456 which are \HII\ regions with embedded stellar associations \citep{bica00}. Region~1 has nine SNRs and high Pearson correlation values for \HI\ with $60\mu m$, $100\mu m,$ and $ 160\mu m$. The excellent wavelet cross correlation of the 21~cm RC with $60~\mu m$ extends down to 50~pc. Region~4 has excellent Pearson correlation of CO with IR emissions at $60~\mu m$, $100~\mu m$ and $160~\mu m$ with the correlation increasing as the temperature of the dust emission decreases. Interestingly, the 21~cm flux for Region~1 is only slightly below Region~3 but Region~4 has the second lowest flux at about 1/2 of the flux of Region~1. The CO data in Region~4 contains NGC 456 but much of the remainder of the region is not covered in the CO data.

Region~5 has a WLS slope of 0.72 and a thermal fraction of 0.76. The region contains the open cluster NGC\,419 which has been examined by \cite{rube10} who suggests that the star formation history of the cluster displays continuous star formation over 1.2 to 1.9~Gyr. This region exhibits no strong Pearson correlations in any data set pairs investigated here (Table~\ref{tbl:PearsonCorr}). There is excellent wavelet correlation down to 130~pc which is the largest measured by this study of the SMC. Oddly, the Spitzer $70~\mu m$ wavelet correlation with 6~cm RC gives a much smaller correlation threshold, 50~pc, but the Spitzer $70~\mu m$ is consistent with the $60~\mu m$ correlation threshold when correlated to 6~cm and 3~cm RC data.

Region~2 has the smallest WLS slope of the regions with a value of 0.62. It also displays the smallest thermal fraction of 0.54. It contains NGC\,330 which is an open cluster and two SNRs. It contains no strong Pearson correlations. The wavelet correlation threshold is at the resolution limit of 35~pc. The higher resolution wavelet correlation graphs, Figure~\ref{fig:wave15h}, show the the correlation actually goes down to 19~pc.

The thermal fraction of the SMC is estimated here to be 0.43. This is higher than the typical value of $ \sim 0.1$ given by \cite{cond92} for spiral galaxies. It is very close to the value of 0.45 for the LMC measured by \cite{hugh06}. The average value of the thermal fraction of the 5 regions is 0.78 over an area of 2.9 ${\deg ^2}$ whereas the SMC measurement is over 9 ${\deg ^2}$. This suggests that the regions, primarily in the bar, emit mostly thermal RC energy but that the larger area contains significant non-thermal emission from a large population of ${\rm{CR}}{{\rm{e}}^ - }$s that have diffused out of the bar. The star formation activity is in the bar where the \HI\ has high density. The non-thermal emission is taking place where the \HI\ has low density outside of the bar. The small size of the SMC suggests that the cosmic ray leakage into the nearby areas would be substantial. The total 21~cm flux from the regions is 26.5~Jy and the total flux in the ${3^ \circ } \times {3^ \circ }$ data set used for these thermal fraction calculations is 36~Jy.

It is interesting to note that the q values that are calculated including only the thermal portion of the RC emission satisfies Y01's range of q values for normal galaxies for the five regions: 1)~2.90, 2)~2.85, 3)~2.72, 4)~2.83, 5)~2.85.  The $''$greater$''$  SMC has a q value of 3.02 which is just at the  98\% q value consistent with Y01's correlation study.
\section{Conclusions}
 \label{s:Conclusions}

We show in this study that the FIR/RC correlations for the SMC and for each of the five regions defined here, are consistent with Y01's regression from 1750 galaxies. The study presented here considers the $''$greater$''$ SMC and the sub-regions of the bar and wing as well.

Pearson correlation coefficients of RC 21~cm emissions with $60~\mu$m and $100~\mu$m emissions have positive correlation only in Region~3 which is the region with the highest star~formation rate. Not surprisingly, Region~3 also has positive Pearson correlation of $H\alpha$ with RC and FIR emissions. Interestingly, Region~1 has twice as much \HI\ and $H_2$ as Region~3 and has the second highest star formation rate in the SMC. Its strongest Pearson correlations are in terms of \HI\ with FIR emissions. Since the star formation rate in the SMC is predicted to be stochastic, Region~1 may have a suppressed supernova rate at this time.

Pixel by pixel scatter plots indicate two populations of correlations in terms of 21~cm RC and $60~\mu$m emissions. The highest slope is from Region~3 which also has the largest difference between the WLS and OLS slope values, 1.14-0.98, 14.5\%. Region~1 has the second largest slope and the second largest difference of slopes of 0.85-0.73, 13.7\%. This correlates to the regions with the strongest star formation rate having the strongest indication of two distinct emission populations. The other regions and the SMC have smaller slopes and less significant differences of 7.8\% to 3.6\%.

The wavelet cross correlation scale factors for the FIR with the RC for the regions are all within 20 to 60~pc except for Region~5 which is $\approx$130~pc and may have few features at small scale factors. There is no clear trend displayed in this correlation. It is interesting to note that the region with the highest thermal fraction, Region~3, and the region with the lowest thermal fraction, Region~2, have the the smallest wavelet correlation scale factors. H$\alpha$ has wavelet cross correlation scale factors with the 6~cm RC emission of between 9 and 14~pc for all of the regions except for Region~5 which is 87~pc.

The regions have q values of: 1)~2.80, 2)~2.58, 3)~2.70, 4)~2.75, 5)~2.73. The measured value of q for the SMC is 2.65 which includes portions of the greater SMC that are not included in regions comprising the bar and wing. The SMC and all of the regions have q values above Y01's correlation value of 2.34 but well within the values of 1.64 to 3.04 between which 98\% of the q values for the galaxies measured by Y01 were found.  This suggests that the RC emission in the SMC is slightly deficient compared to the FIR emission.  It could be due to ${\rm{CR}}{{\rm{e}}^ - }$s escaping the bar or wing of the SMC without losing much of their energy. Since the non-thermal fraction of the RC emission is so small, the non-thermal RC emission is significantly under-luminous considering that the non-thermal fraction is expected to be dominant. This may be due to the small physical size of the SMC and a low magnetic field density.

The SMC displays a thermal fraction that is much higher than is common for disk galaxies. Disk galaxies typically have a thermal fraction of 0.1 \citep{cond92} while the SMC is measured here
to have a value of 0.43. The regions have even higher measured values of the thermal fraction, 1)~0.81, 2)~0.54, 3)~0.95, 4)~0.84, and 5)~0.76.  This suggests that galaxies can have most of their RC radiation derived from thermal sources powered by the radiation field from high mass stars in \HII\ regions rather than by SNRs. There appears to be a weak correlation between the thermal fraction and the q values.  The thermal fraction values above 0.5 correspond to q values above 2.7 and thermal fraction values below about 0.5 have q values of below 2.7.  The SMC and all five of the defined regions have FIR/RC correlations consistent with Y01 q values when considering only thermal RC radiation. Thermal radiation can be the major source of RC emission from a galaxy suggesting that measurements of the RC emission of galaxies could lead to over estimating the super nova rate by assuming the 0.1 ratio of thermal to non thermal radiation.

%
\acknowledgments

We would like to thank the anonymous referee for a careful reading this paper and for the astute comments that improved the final version of this paper significantly.

\bibliographystyle{spr-mp-nameyear-cnd}        
\bibliography{hleverenz_ref}                   

\begin{thebibliography}{72}
\ifx \bisbn   \undefined \def \bisbn  #1{ISBN #1}\fi
\ifx \binits  \undefined \def \binits#1{#1} \fi
\ifx \bauthor  \undefined \def \bauthor#1{#1} \fi
\ifx \batitle  \undefined \def \batitle#1{#1} \fi
\ifx \bjtitle  \undefined \def \bjtitle#1{#1}\fi
\ifx \bvolume  \undefined \def \bvolume#1{\textbf{#1}}\fi
\ifx \byear  \undefined \def \byear#1{#1} \fi
\ifx \bissue  \undefined \def \bissue#1{#1} \fi
\ifx \bfpage  \undefined \def \bfpage#1{#1} \fi
\ifx \blpage  \undefined \def \blpage #1{#1} \fi
\ifx \burl  \undefined \def \burl#1{\textsf{#1}} \fi
\ifx \doiurl  \undefined \def \doiurl#1{\textsf{#1}} \fi
\ifx \betal  \undefined \def \betal{\textit{et al.}} \fi
\ifx \binstitute  \undefined \def \binstitute#1{#1} \fi
\ifx \bctitle  \undefined \def \bctitle#1{#1} \fi
\ifx \beditor  \undefined \def \beditor#1{#1} \fi
\ifx \bpublisher  \undefined \def \bpublisher#1{#1} \fi
\ifx \bbtitle  \undefined \def \bbtitle#1{#1} \fi
\ifx \bedition  \undefined \def \bedition#1{#1} \fi
\ifx \bseriesno  \undefined \def \bseriesno#1{#1} \fi
\ifx \blocation  \undefined \def \blocation#1{#1} \fi
\ifx \bsertitle  \undefined \def \bsertitle#1{#1} \fi
\ifx \bsnm \undefined \def \bsnm#1{#1} \fi
\ifx \bsuffix \undefined \def \bsuffix#1{#1} \fi
\ifx \bparticle \undefined \def \bparticle#1{#1} \fi
\ifx \barticle \undefined \def \barticle#1{#1} \fi
\ifx \botherref \undefined \def \botherref #1{#1} \fi
\ifx \url \undefined \def \url#1{\textsf{#1}} \fi
\ifx \bchapter \undefined \def \bchapter#1{#1} \fi
\ifx \bbook \undefined \def \bbook#1{#1} \fi
\ifx \bcomment \undefined \def \bcomment#1{#1} \fi
\ifx \oauthor \undefined \def \oauthor#1{#1} \fi
\ifx \citeauthoryear \undefined \def \citeauthoryear#1{#1} \fi
\def \endbibitem {}

\bibitem[\protect\citeauthoryear{Appleton \textit{et~al.}}{2004}]{appl04}
\begin{barticle}
\bauthor{\bsnm{Appleton}, \binits{P.N.}}, \bauthor{\bsnm{Fadda},
  \binits{D.T.}}, \bauthor{\bsnm{Marleau}, \binits{F.R.}},
  \bauthor{\bsnm{Frayer}, \binits{D.T.}}, \bauthor{\bsnm{Helou}, \binits{G.}},
  \bauthor{\bsnm{Condon}, \binits{J.J.}}, \bauthor{\bsnm{Choi}, \binits{P.I.}},
  \bauthor{\bsnm{Yan}, \binits{L.}}, \bauthor{\bsnm{Lacy}, \binits{M.}},
  \bauthor{\bsnm{Wilson}, \binits{G.}}, \bauthor{\bsnm{Armus}, \binits{L.}},
  \bauthor{\bsnm{Chapman}, \binits{S.C.}}, \bauthor{\bsnm{Fang}, \binits{F.}},
  \bauthor{\bsnm{Heinrichson}, \binits{I.}}, \bauthor{\bsnm{Im}, \binits{M.}},
  \bauthor{\bsnm{Jannuzi}, \binits{B.T.}}, \bauthor{\bsnm{Storrie-Lombardi},
  \binits{L.J.}}, \bauthor{\bsnm{Shupe}, \binits{D.}}, \bauthor{\bsnm{Soifer},
  \binits{B.T.}}, \bauthor{\bsnm{Squires}, \binits{G.}},
  \bauthor{\bsnm{Teplitz}, \binits{H.I.}}:
\bjtitle{The Astrophysical Journal Supplement Series}
\bvolume{154},
\bfpage{147}
(\byear{2004})
\end{barticle}
\endbibitem

\bibitem[\protect\citeauthoryear{Beck}{2000}]{beck00}
\begin{barticle}
\bauthor{\bsnm{Beck}, \binits{R.}}:
\bjtitle{Philosophical Transactions: Mathematical, Physical and Engineering
  Sciences}
\bvolume{358}(\bissue{1767}),
\bfpage{777}
(\byear{2000}).
\bcomment{Hand generated referenc.}
\end{barticle}
\endbibitem

\bibitem[\protect\citeauthoryear{Bica and Dutra}{2000}]{bica00}
\begin{barticle}
\bauthor{\bsnm{Bica}, \binits{E.}}, \bauthor{\bsnm{Dutra}, \binits{C.M.}}:
\bjtitle{The Astronomical Journal}
\bvolume{119},
\bfpage{1214}
(\byear{2000})
\end{barticle}
\endbibitem

\bibitem[\protect\citeauthoryear{Blackburn, Payne, and Hayes}{1995}]{blac95}
\begin{botherref}
\oauthor{\bsnm{Blackburn}, \binits{J.K.}}, \oauthor{\bsnm{Payne},
  \binits{H.E.}}, \oauthor{\bsnm{Hayes}, \binits{J.J.E.}}:
1995
Ftools: A fits data processing and analysis software package
\end{botherref}
\endbibitem

\bibitem[\protect\citeauthoryear{Bolatto \textit{et~al.}}{2007}]{bola07}
\begin{barticle}
\bauthor{\bsnm{Bolatto}, \binits{A.D.}}, \bauthor{\bsnm{Simon}, \binits{J.D.}},
  \bauthor{\bsnm{Stanimirovic}, \binits{S.}}, \bauthor{\bparticle{van
  }\bsnm{Loon}, \binits{J.T.}}, \bauthor{\bsnm{Shah}, \binits{R.Y.}},
  \bauthor{\bsnm{Venn}, \binits{K.}}, \bauthor{\bsnm{Leroy}, \binits{A.K.}},
  \bauthor{\bsnm{Sandstrom}, \binits{K.}}, \bauthor{\bsnm{Jackson},
  \binits{J.M.}}, \bauthor{\bsnm{Israel}, \binits{F.P.}}, \bauthor{\bsnm{Li},
  \binits{A.}}, \bauthor{\bsnm{Staveley-Smith}, \binits{L.}},
  \bauthor{\bsnm{Bot}, \binits{C.}}, \bauthor{\bsnm{Boulanger}, \binits{F.}},
  \bauthor{\bsnm{Rubio}, \binits{M.}}:
\bjtitle{The Astrophysical Journal}
\bvolume{655},
\bfpage{212}
(\byear{2007})
\end{barticle}
\endbibitem

\bibitem[\protect\citeauthoryear{Bot \textit{et~al.}}{2010}]{bot10}
\begin{barticle}
\bauthor{\bsnm{Bot}, \binits{C.}}, \bauthor{\bsnm{Ysard}, \binits{N.}},
  \bauthor{\bsnm{Paradis}, \binits{D.}}, \bauthor{\bsnm{Bernard},
  \binits{J.P.}}, \bauthor{\bsnm{Lagache}, \binits{G.}},
  \bauthor{\bsnm{Israel}, \binits{F.P.}}, \bauthor{\bsnm{Wall}, \binits{W.F.}}:
\bjtitle{Astronomy and Astrophysics}
\bvolume{523},
\bfpage{20}
(\byear{2010})
\end{barticle}
\endbibitem

\bibitem[\protect\citeauthoryear{Cignoni \textit{et~al.}}{2011}]{cign11}
\begin{barticle}
\bauthor{\bsnm{Cignoni}, \binits{M.}}, \bauthor{\bsnm{Tosi}, \binits{M.}},
  \bauthor{\bsnm{Sabbi}, \binits{E.}}, \bauthor{\bsnm{Nota}, \binits{A.}},
  \bauthor{\bsnm{Gallagher}, \binits{J.S.}}:
\bjtitle{The Astronomical Journal}
\bvolume{141},
\bfpage{31}
(\byear{2011})
\end{barticle}
\endbibitem

\bibitem[\protect\citeauthoryear{Cioni, Habing, and Israel}{2000}]{cion00}
\begin{barticle}
\bauthor{\bsnm{Cioni}, \binits{M.R.L.}}, \bauthor{\bsnm{Habing},
  \binits{H.J.}}, \bauthor{\bsnm{Israel}, \binits{F.P.}}:
\bjtitle{Astronomy and Astrophysics}
\bvolume{358},
\bfpage{9}
(\byear{2000})
\end{barticle}
\endbibitem

\bibitem[\protect\citeauthoryear{Condon}{1992}]{cond92}
\begin{barticle}
\bauthor{\bsnm{Condon}, \binits{J.J.}}:
\bjtitle{Annual Review of Astronomy and Astrophysics}
\bvolume{30},
\bfpage{575}
(\byear{1992})
\end{barticle}
\endbibitem

\bibitem[\protect\citeauthoryear{Costa \textit{et~al.}}{2011}]{cost11}
\begin{barticle}
\bauthor{\bsnm{Costa}, \binits{E.}}, \bauthor{\bsnm{Méndez}, \binits{R.A.}},
  \bauthor{\bsnm{Pedreros}, \binits{M.H.}}, \bauthor{\bsnm{Moyano},
  \binits{M.}}, \bauthor{\bsnm{Gallart}, \binits{C.}}, \bauthor{\bsnm{Noël},
  \binits{N.}}:
\bjtitle{The Astronomical Journal}
\bvolume{141},
\bfpage{136}
(\byear{2011})
\end{barticle}
\endbibitem

\bibitem[\protect\citeauthoryear{Crawford \textit{et~al.}}{2011}]{craw11}
\begin{botherref}
\oauthor{\bsnm{Crawford}, \binits{E.J.}}, \oauthor{\bsnm{Filipovic},
  \binits{M.D.}}, \oauthor{\bsnm{De~Horta}, \binits{A.Y.}},
  \oauthor{\bsnm{Wong}, \binits{G.F.}}, \oauthor{\bsnm{Tothill},
  \binits{N.F.H.}}, \oauthor{\bsnm{Draskovic}, \binits{D.}},
  \oauthor{\bsnm{Collier}, \binits{J.D.}}, \oauthor{\bsnm{Galvin},
  \binits{T.J.}}:
2011
New 6 and 3-cm radio-continuum maps of the small magellanic cloud. part i - the
  maps.
10 Pages, 6 figures, accepted for publication in the Serbian Astronomical
  Journal
\end{botherref}
\endbibitem

\bibitem[\protect\citeauthoryear{Dickel \textit{et~al.}}{2010}]{dick10}
\begin{barticle}
\bauthor{\bsnm{Dickel}, \binits{J.R.}}, \bauthor{\bsnm{Gruendl},
  \binits{R.A.}}, \bauthor{\bsnm{McIntyre}, \binits{V.J.}},
  \bauthor{\bsnm{Amy}, \binits{S.W.}}:
\bjtitle{The Astronomical Journal}
\bvolume{140},
\bfpage{1511}
(\byear{2010})
\end{barticle}
\endbibitem

\bibitem[\protect\citeauthoryear{Diehl \textit{et~al.}}{2006}]{dieh06}
\begin{barticle}
\bauthor{\bsnm{Diehl}, \binits{R.}}, \bauthor{\bsnm{Halloin}, \binits{H.}},
  \bauthor{\bsnm{Kretschmer}, \binits{K.}}, \bauthor{\bsnm{Lichti},
  \binits{G.G.}}, \bauthor{\bsnm{Schönfelder}, \binits{V.}},
  \bauthor{\bsnm{Strong}, \binits{A.W.}}, \bauthor{\bparticle{von
  }\bsnm{Kienlin}, \binits{A.}}, \bauthor{\bsnm{Wang}, \binits{W.}},
  \bauthor{\bsnm{Jean}, \binits{P.}}, \bauthor{\bsnm{Knödlseder},
  \binits{J.}}, \bauthor{\bsnm{Roques}, \binits{J.P.}},
  \bauthor{\bsnm{Weidenspointner}, \binits{G.}}, \bauthor{\bsnm{Schanne},
  \binits{S.}}, \bauthor{\bsnm{Hartmann}, \binits{D.H.}},
  \bauthor{\bsnm{Winkler}, \binits{C.}}, \bauthor{\bsnm{Wunderer},
  \binits{C.}}:
\bjtitle{Nature}
\bvolume{439},
\bfpage{45}
(\byear{2006})
\end{barticle}
\endbibitem

\bibitem[\protect\citeauthoryear{Filipovic \textit{et~al.}}{1995}]{fili95}
\begin{barticle}
\bauthor{\bsnm{Filipovic}, \binits{M.D.}}, \bauthor{\bsnm{Haynes},
  \binits{R.F.}}, \bauthor{\bsnm{White}, \binits{G.L.}}, \bauthor{\bsnm{Jones},
  \binits{P.A.}}, \bauthor{\bsnm{Klein}, \binits{U.}},
  \bauthor{\bsnm{Wielebinski}, \binits{R.}}:
\bjtitle{Astronomy and Astrophysics Supplement Series}
\bvolume{111},
\bfpage{311}
(\byear{1995})
\end{barticle}
\endbibitem

\bibitem[\protect\citeauthoryear{Filipovic \textit{et~al.}}{1997}]{fili97}
\begin{barticle}
\bauthor{\bsnm{Filipovic}, \binits{M.D.}}, \bauthor{\bsnm{Jones},
  \binits{P.A.}}, \bauthor{\bsnm{White}, \binits{G.L.}},
  \bauthor{\bsnm{Haynes}, \binits{R.F.}}, \bauthor{\bsnm{Klein}, \binits{U.}},
  \bauthor{\bsnm{Wielebinski}, \binits{R.}}:
\bjtitle{Astronomy and Astrophysics Supplement Series}
\bvolume{121},
\bfpage{321}
(\byear{1997})
\end{barticle}
\endbibitem

\bibitem[\protect\citeauthoryear{Filipovic \textit{et~al.}}{1998}]{fili98}
\begin{barticle}
\bauthor{\bsnm{Filipovic}, \binits{M.D.}}, \bauthor{\bsnm{Haynes},
  \binits{R.F.}}, \bauthor{\bsnm{White}, \binits{G.L.}}, \bauthor{\bsnm{Jones},
  \binits{P.A.}}:
\bjtitle{Astronomy and Astrophysics Supplement Series}
\bvolume{130},
\bfpage{421}
(\byear{1998})
\end{barticle}
\endbibitem

\bibitem[\protect\citeauthoryear{Filipovic \textit{et~al.}}{1998a}]{fili98a}
\begin{barticle}
\bauthor{\bsnm{Filipovic}, \binits{M.D.}}, \bauthor{\bsnm{Jones},
  \binits{P.A.}}, \bauthor{\bsnm{White}, \binits{G.L.}},
  \bauthor{\bsnm{Haynes}, \binits{R.F.}}:
\bjtitle{Astronomy and Astrophysics Supplement Series}
\bvolume{130},
\bfpage{441}
(\byear{1998}a)
\end{barticle}
\endbibitem

\bibitem[\protect\citeauthoryear{Filipovic \textit{et~al.}}{1998b}]{fili98b}
\begin{barticle}
\bauthor{\bsnm{Filipovic}, \binits{M.D.}}, \bauthor{\bsnm{Pietsch},
  \binits{W.}}, \bauthor{\bsnm{Haynes}, \binits{R.F.}}, \bauthor{\bsnm{White},
  \binits{G.L.}}, \bauthor{\bsnm{Jones}, \binits{P.A.}},
  \bauthor{\bsnm{Wielebinski}, \binits{R.}}, \bauthor{\bsnm{Klein},
  \binits{U.}}, \bauthor{\bsnm{Dennerl}, \binits{K.}}, \bauthor{\bsnm{Kahabka},
  \binits{P.}}, \bauthor{\bsnm{Lazendic}, \binits{J.S.}}:
\bjtitle{Astronomy and Astrophysics Supplement Series}
\bvolume{127},
\bfpage{119}
(\byear{1998}b)
\end{barticle}
\endbibitem

\bibitem[\protect\citeauthoryear{Filipovic \textit{et~al.}}{2002}]{fili02}
\begin{barticle}
\bauthor{\bsnm{Filipovic}, \binits{M.D.}}, \bauthor{\bsnm{Bohlsen},
  \binits{T.}}, \bauthor{\bsnm{Reid}, \binits{W.}},
  \bauthor{\bsnm{Staveley-Smith}, \binits{L.}}, \bauthor{\bsnm{Jones},
  \binits{P.A.}}, \bauthor{\bsnm{Nohejl}, \binits{K.}},
  \bauthor{\bsnm{Goldstein}, \binits{G.}}:
\bjtitle{Monthly Notices of the Royal Astronomical Society}
\bvolume{335},
\bfpage{1085}
(\byear{2002}).
\bcomment{DOI: 10.1046/j.1365-8711.2002.05702.x}
\end{barticle}
\endbibitem

\bibitem[\protect\citeauthoryear{Filipovic \textit{et~al.}}{2005}]{fili05}
\begin{barticle}
\bauthor{\bsnm{Filipovic}, \binits{M.D.}}, \bauthor{\bsnm{Payne},
  \binits{J.L.}}, \bauthor{\bsnm{Reid}, \binits{W.}}, \bauthor{\bsnm{Danforth},
  \binits{C.W.}}, \bauthor{\bsnm{Staveley-Smith}, \binits{L.}},
  \bauthor{\bsnm{Jones}, \binits{P.A.}}, \bauthor{\bsnm{White}, \binits{G.L.}}:
\bjtitle{Monthly Notices of the Royal Astronomical Society}
\bvolume{364},
\bfpage{217}
(\byear{2005}).
\bcomment{DOI: 10.1111/j.1365-2966.2005.09554.x}
\end{barticle}
\endbibitem

\bibitem[\protect\citeauthoryear{Frick \textit{et~al.}}{2001}]{fric01}
\begin{barticle}
\bauthor{\bsnm{Frick}, \binits{P.}}, \bauthor{\bsnm{Beck}, \binits{R.}},
  \bauthor{\bsnm{Berkhuijsen}, \binits{E.M.}}, \bauthor{\bsnm{Patrickeyev},
  \binits{I.}}:
\bjtitle{Monthly Notices of the Royal Astronomical Society}
\bvolume{327},
\bfpage{1145}
(\byear{2001}).
\bcomment{DOI: 10.1046/j.1365-8711.2001.04812.x; eprintid:
  arXiv:astro-ph/0109017}
\end{barticle}
\endbibitem

\bibitem[\protect\citeauthoryear{Gooch and Barnes}{1996}]{gooc96}
\begin{botherref}
\oauthor{\bsnm{Gooch}, \binits{R.}}, \oauthor{\bsnm{Barnes}, \binits{J.}}:
1996
Karma: a visualization test-bed
\end{botherref}
\endbibitem

\bibitem[\protect\citeauthoryear{Haslam and Osborne}{1987}]{hasl87}
\begin{barticle}
\bauthor{\bsnm{Haslam}, \binits{C.G.T.}}, \bauthor{\bsnm{Osborne},
  \binits{J.L.}}:
\bjtitle{Nature}
\bvolume{327},
\bfpage{211}
(\byear{1987})
\end{barticle}
\endbibitem

\bibitem[\protect\citeauthoryear{Haynes \textit{et~al.}}{1991}]{hayn91}
\begin{barticle}
\bauthor{\bsnm{Haynes}, \binits{R.F.}}, \bauthor{\bsnm{Klein}, \binits{U.}},
  \bauthor{\bsnm{Wayte}, \binits{S.R.}}, \bauthor{\bsnm{Wielebinski},
  \binits{R.}}, \bauthor{\bsnm{Murray}, \binits{J.D.}}, \bauthor{\bsnm{Bajaja},
  \binits{E.}}, \bauthor{\bsnm{Meinert}, \binits{D.}},
  \bauthor{\bsnm{Buczilowski}, \binits{U.R.}}, \bauthor{\bsnm{Harnett},
  \binits{J.I.}}, \bauthor{\bsnm{Hunt}, \binits{A.J.}}, \bauthor{\bsnm{Wark},
  \binits{R.}}, \bauthor{\bsnm{Sciacca}, \binits{L.}}:
\bjtitle{Astronomy and Astrophysics}
\bvolume{252},
\bfpage{475}
(\byear{1991})
\end{barticle}
\endbibitem

\bibitem[\protect\citeauthoryear{Helder \textit{et~al.}}{2009}]{held09}
\begin{barticle}
\bauthor{\bsnm{Helder}, \binits{E.A.}}, \bauthor{\bsnm{Vink}, \binits{J.}},
  \bauthor{\bsnm{Bassa}, \binits{C.G.}}, \bauthor{\bsnm{Bamba}, \binits{A.}},
  \bauthor{\bsnm{Bleeker}, \binits{J.A.M.}}, \bauthor{\bsnm{Funk},
  \binits{S.}}, \bauthor{\bsnm{Ghavamian}, \binits{P.}},
  \bauthor{\bparticle{van~der }\bsnm{Heyden}, \binits{K.J.}},
  \bauthor{\bsnm{Verbunt}, \binits{F.}}, \bauthor{\bsnm{Yamazaki},
  \binits{R.}}:
\bjtitle{Science}
\bvolume{325},
\bfpage{719}
(\byear{2009}).
\bcomment{DOI: 10.1126/science.1173383; eprintid: arXiv:0906.4553}
\end{barticle}
\endbibitem

\bibitem[\protect\citeauthoryear{Helou and Bicay}{1993}]{helo93}
\begin{barticle}
\bauthor{\bsnm{Helou}, \binits{G.}}, \bauthor{\bsnm{Bicay}, \binits{M.D.}}:
\bjtitle{Astrophysical Journal}
\bvolume{415},
\bfpage{93}
(\byear{1993})
\end{barticle}
\endbibitem

\bibitem[\protect\citeauthoryear{Helou \textit{et~al.}}{1988}]{helo88}
\begin{barticle}
\bauthor{\bsnm{Helou}, \binits{G.}}, \bauthor{\bsnm{Khan}, \binits{I.R.}},
  \bauthor{\bsnm{Malek}, \binits{L.}}, \bauthor{\bsnm{Boehmer}, \binits{L.}}:
\bjtitle{Astrophysical Journal Supplement Series}
\bvolume{68},
\bfpage{151}
(\byear{1988})
\end{barticle}
\endbibitem

\bibitem[\protect\citeauthoryear{Hilditch, Howarth, and Harries}{2005}]{hild05}
\begin{barticle}
\bauthor{\bsnm{Hilditch}, \binits{R.W.}}, \bauthor{\bsnm{Howarth},
  \binits{I.D.}}, \bauthor{\bsnm{Harries}, \binits{T.J.}}:
\bjtitle{Monthly Notices of the Royal Astronomical Society}
\bvolume{357},
\bfpage{304}
(\byear{2005})
\end{barticle}
\endbibitem

\bibitem[\protect\citeauthoryear{Hoernes, Berkhuijsen, and Xu}{1998}]{hoer98}
\begin{barticle}
\bauthor{\bsnm{Hoernes}, \binits{P.}}, \bauthor{\bsnm{Berkhuijsen},
  \binits{E.M.}}, \bauthor{\bsnm{Xu}, \binits{C.}}:
\bjtitle{Astronomy and Astrophysics}
\bvolume{334},
\bfpage{57}
(\byear{1998})
\end{barticle}
\endbibitem

\bibitem[\protect\citeauthoryear{Hughes \textit{et~al.}}{2006}]{hugh06}
\begin{barticle}
\bauthor{\bsnm{Hughes}, \binits{A.}}, \bauthor{\bsnm{Wong}, \binits{T.}},
  \bauthor{\bsnm{Ekers}, \binits{R.}}, \bauthor{\bsnm{Staveley-Smith},
  \binits{L.}}, \bauthor{\bsnm{Filipovic}, \binits{M.}},
  \bauthor{\bsnm{Maddison}, \binits{S.}}, \bauthor{\bsnm{Fukui}, \binits{Y.}},
  \bauthor{\bsnm{Mizuno}, \binits{N.}}:
\bjtitle{Monthly Notices of the Royal Astronomical Society}
\bvolume{370},
\bfpage{363}
(\byear{2006})
\end{barticle}
\endbibitem

\bibitem[\protect\citeauthoryear{Israel \textit{et~al.}}{2010}]{isra10}
\begin{barticle}
\bauthor{\bsnm{Israel}, \binits{F.P.}}, \bauthor{\bsnm{Wall}, \binits{W.F.}},
  \bauthor{\bsnm{Raban}, \binits{D.}}, \bauthor{\bsnm{Reach}, \binits{W.T.}},
  \bauthor{\bsnm{Bot}, \binits{C.}}, \bauthor{\bsnm{Oonk}, \binits{J.B.R.}},
  \bauthor{\bsnm{Ysard}, \binits{N.}}, \bauthor{\bsnm{Bernard}, \binits{J.P.}}:
\bjtitle{Astronomy and Astrophysics}
\bvolume{519},
\bfpage{67}
(\byear{2010})
\end{barticle}
\endbibitem

\bibitem[\protect\citeauthoryear{Joye \textit{et~al.}}{2003}]{joye03}
\begin{botherref}
\oauthor{\bsnm{Joye}, \binits{W.A.}}, \oauthor{\bsnm{Mandel}, \binits{E.}},
  \oauthor{\bsnm{Jedrzejewski}, \binits{R.I.}}, \oauthor{\bsnm{Hook},
  \binits{R.N.}}:
2003
New features of saoimage ds9
\end{botherref}
\endbibitem

\bibitem[\protect\citeauthoryear{Klein and Graeve}{1986}]{klein86}
\begin{barticle}
\bauthor{\bsnm{Klein}, \binits{U.}}, \bauthor{\bsnm{Graeve}, \binits{R.}}:
\bjtitle{Astronomy and Astrophysics}
\bvolume{161},
\bfpage{155}
(\byear{1986})
\end{barticle}
\endbibitem

\bibitem[\protect\citeauthoryear{Klein, Grave, and Beck}{1984}]{klein84}
\begin{botherref}
\oauthor{\bsnm{Klein}, \binits{U.}}, \oauthor{\bsnm{Grave}, \binits{R.}},
  \oauthor{\bsnm{Beck}, \binits{R.}}:
1984
Radio continuum emission from magellanic-type and dwarf irregular galaxies
\end{botherref}
\endbibitem

\bibitem[\protect\citeauthoryear{Klein \textit{et~al.}}{1993}]{klein93}
\begin{barticle}
\bauthor{\bsnm{Klein}, \binits{U.}}, \bauthor{\bsnm{Haynes}, \binits{R.F.}},
  \bauthor{\bsnm{Wielebinski}, \binits{R.}}, \bauthor{\bsnm{Meinert},
  \binits{D.}}:
\bjtitle{Astronomy and Astrophysics}
\bvolume{271},
\bfpage{402}
(\byear{1993})
\end{barticle}
\endbibitem

\bibitem[\protect\citeauthoryear{Koornneef and de~Boer}{1984}]{koor84}
\begin{botherref}
\oauthor{\bsnm{Koornneef}, \binits{J.}}, \oauthor{\bparticle{de~}\bsnm{Boer},
  \binits{K.S.D.}}:
1984
Gas-to-dust ratios in the magellanic clouds
\end{botherref}
\endbibitem

\bibitem[\protect\citeauthoryear{Leroy \textit{et~al.}}{2011}]{lero11}
\begin{barticle}
\bauthor{\bsnm{Leroy}, \binits{A.K.}}, \bauthor{\bsnm{Bolatto}, \binits{A.}},
  \bauthor{\bsnm{Gordon}, \binits{K.}}, \bauthor{\bsnm{Sandstrom},
  \binits{K.}}, \bauthor{\bsnm{Gratier}, \binits{P.}},
  \bauthor{\bsnm{Rosolowsky}, \binits{E.}}, \bauthor{\bsnm{Engelbracht},
  \binits{C.W.}}, \bauthor{\bsnm{Mizuno}, \binits{N.}},
  \bauthor{\bsnm{Corbelli}, \binits{E.}}, \bauthor{\bsnm{Fukui}, \binits{Y.}},
  \bauthor{\bsnm{Kawamura}, \binits{A.}}:
\bjtitle{The Astrophysical Journal}
\bvolume{737},
\bfpage{12}
(\byear{2011})
\end{barticle}
\endbibitem

\bibitem[\protect\citeauthoryear{Leroy \textit{et~al.}}{2007}]{lero07}
\begin{barticle}
\bauthor{\bsnm{Leroy}, \binits{A.}}, \bauthor{\bsnm{Bolatto}, \binits{A.}},
  \bauthor{\bsnm{Stanimirovic}, \binits{S.}}, \bauthor{\bsnm{Mizuno},
  \binits{N.}}, \bauthor{\bsnm{Israel}, \binits{F.}}, \bauthor{\bsnm{Bot},
  \binits{C.}}:
\bjtitle{Astrophysical Journal}
\bvolume{658},
\bfpage{1027}
(\byear{2007}).
\bcomment{DOI: 10.1086/511150; eprintid: arXiv:astro-ph/0611687}
\end{barticle}
\endbibitem

\bibitem[\protect\citeauthoryear{Loiseau \textit{et~al.}}{1987}]{lois87}
\begin{barticle}
\bauthor{\bsnm{Loiseau}, \binits{N.}}, \bauthor{\bsnm{Klein}, \binits{U.}},
  \bauthor{\bsnm{Greybe}, \binits{A.}}, \bauthor{\bsnm{Wielebinski},
  \binits{R.}}, \bauthor{\bsnm{Haynes}, \binits{R.F.}}:
\bjtitle{Astronomy and Astrophysics}
\bvolume{178},
\bfpage{62}
(\byear{1987})
\end{barticle}
\endbibitem

\bibitem[\protect\citeauthoryear{Maragoudaki \textit{et~al.}}{2001}]{mara01}
\begin{barticle}
\bauthor{\bsnm{Maragoudaki}, \binits{F.}}, \bauthor{\bsnm{Kontizas},
  \binits{M.}}, \bauthor{\bsnm{Morgan}, \binits{D.H.}},
  \bauthor{\bsnm{Kontizas}, \binits{E.}}, \bauthor{\bsnm{Dapergolas},
  \binits{A.}}, \bauthor{\bsnm{Livanou}, \binits{E.}}:
\bjtitle{Astronomy and Astrophysics}
\bvolume{379},
\bfpage{864}
(\byear{2001})
\end{barticle}
\endbibitem

\bibitem[\protect\citeauthoryear{Martin, Maurice, and Lequeux}{1989}]{mart89}
\begin{barticle}
\bauthor{\bsnm{Martin}, \binits{N.}}, \bauthor{\bsnm{Maurice}, \binits{E.}},
  \bauthor{\bsnm{Lequeux}, \binits{J.}}:
\bjtitle{Astronomy and Astrophysics}
\bvolume{215},
\bfpage{219}
(\byear{1989})
\end{barticle}
\endbibitem

\bibitem[\protect\citeauthoryear{Matlab}{2010}]{matl10}
\begin{botherref}
\oauthor{\bsnm{Matlab}}:
2010
Matlab 7.10.0,
The MathWorks Inc.
\end{botherref}
\endbibitem

\bibitem[\protect\citeauthoryear{Miville-Deschênes and Lagache}{2005}]{mivi05}
\begin{barticle}
\bauthor{\bsnm{Miville-Deschênes}, \binits{M.A.}}, \bauthor{\bsnm{Lagache},
  \binits{G.}}:
\bjtitle{The Astrophysical Journal Supplement Series}
\bvolume{157},
\bfpage{302}
(\byear{2005})
\end{barticle}
\endbibitem

\bibitem[\protect\citeauthoryear{Mizuno \textit{et~al.}}{2001}]{mizu01}
\begin{barticle}
\bauthor{\bsnm{Mizuno}, \binits{N.}}, \bauthor{\bsnm{Rubio}, \binits{M.}},
  \bauthor{\bsnm{Mizuno}, \binits{A.}}, \bauthor{\bsnm{Yamaguchi},
  \binits{R.}}, \bauthor{\bsnm{Onishi}, \binits{T.}}, \bauthor{\bsnm{Fukui},
  \binits{Y.}}:
\bjtitle{Publications of the Astronomical Society of Japan}
\bvolume{53},
\bfpage{45}
(\byear{2001})
\end{barticle}
\endbibitem

\bibitem[\protect\citeauthoryear{Murai and Fujimoto}{1980}]{mura80}
\begin{barticle}
\bauthor{\bsnm{Murai}, \binits{T.}}, \bauthor{\bsnm{Fujimoto}, \binits{M.}}:
\bjtitle{Publications of the Astronomical Society of Japan}
\bvolume{32},
\bfpage{581}
(\byear{1980})
\end{barticle}
\endbibitem

\bibitem[\protect\citeauthoryear{Niklas, Klein, and Wielebinski}{1997}]{nikl97}
\begin{barticle}
\bauthor{\bsnm{Niklas}, \binits{S.}}, \bauthor{\bsnm{Klein}, \binits{U.}},
  \bauthor{\bsnm{Wielebinski}, \binits{R.}}:
\bjtitle{Astronomy and Astrophysics}
\bvolume{322},
\bfpage{19}
(\byear{1997})
\end{barticle}
\endbibitem

\bibitem[\protect\citeauthoryear{Payne \textit{et~al.}}{2004}]{payn04}
\begin{barticle}
\bauthor{\bsnm{Payne}, \binits{J.L.}}, \bauthor{\bsnm{Filipovic},
  \binits{M.D.}}, \bauthor{\bsnm{Reid}, \binits{W.}}, \bauthor{\bsnm{Jones},
  \binits{P.A.}}, \bauthor{\bsnm{Staveley-Smith}, \binits{L.}},
  \bauthor{\bsnm{White}, \binits{G.L.}}:
\bjtitle{Monthly Notices of the Royal Astronomical Society}
\bvolume{355},
\bfpage{44}
(\byear{2004}).
\bcomment{DOI: 10.1111/j.1365-2966.2004.08287.x}
\end{barticle}
\endbibitem

\bibitem[\protect\citeauthoryear{Payne \textit{et~al.}}{2007}]{payn07}
\begin{barticle}
\bauthor{\bsnm{Payne}, \binits{J.L.}}, \bauthor{\bsnm{White}, \binits{G.L.}},
  \bauthor{\bsnm{Filipovic}, \binits{M.D.}}, \bauthor{\bsnm{Pannuti},
  \binits{T.G.}}:
\bjtitle{Monthly Notices of the Royal Astronomical Society}
\bvolume{376},
\bfpage{1793}
(\byear{2007}).
\bcomment{DOI: 10.1111/j.1365-2966.2007.11561.x}
\end{barticle}
\endbibitem

\bibitem[\protect\citeauthoryear{Piatek, Pryor, and Olszewski}{2008}]{piat08}
\begin{barticle}
\bauthor{\bsnm{Piatek}, \binits{S.}}, \bauthor{\bsnm{Pryor}, \binits{C.}},
  \bauthor{\bsnm{Olszewski}, \binits{E.W.}}:
\bjtitle{The Astronomical Journal}
\bvolume{135},
\bfpage{1024}
(\byear{2008})
\end{barticle}
\endbibitem

\bibitem[\protect\citeauthoryear{Reid \textit{et~al.}}{2006}]{reid06}
\begin{barticle}
\bauthor{\bsnm{Reid}, \binits{W.A.}}, \bauthor{\bsnm{Payne}, \binits{J.L.}},
  \bauthor{\bsnm{Filipovic}, \binits{M.D.}}, \bauthor{\bsnm{Danforth},
  \binits{C.W.}}, \bauthor{\bsnm{Jones}, \binits{P.A.}}, \bauthor{\bsnm{White},
  \binits{G.L.}}, \bauthor{\bsnm{Staveley-Smith}, \binits{L.}}:
\bjtitle{Monthly Notices of the Royal Astronomical Society}
\bvolume{367},
\bfpage{1379}
(\byear{2006}).
\bcomment{DOI: 10.1111/j.1365-2966.2006.10017.x}
\end{barticle}
\endbibitem

\bibitem[\protect\citeauthoryear{Rubele, Kerber, and Girardi}{2010}]{rube10}
\begin{barticle}
\bauthor{\bsnm{Rubele}, \binits{S.}}, \bauthor{\bsnm{Kerber}, \binits{L.}},
  \bauthor{\bsnm{Girardi}, \binits{L.}}:
\bjtitle{Monthly Notices of the Royal Astronomical Society}
\bvolume{403},
\bfpage{1156}
(\byear{2010})
\end{barticle}
\endbibitem

\bibitem[\protect\citeauthoryear{Sandage and Tammann}{1981}]{sand81}
\begin{bbook}
\bauthor{\bsnm{Sandage}, \binits{A.}}, \bauthor{\bsnm{Tammann}, \binits{G.A.}}:
\bbtitle{A revised shapley-ames catalog of bright galaxies}.
\bsertitle{Carnegie Institution of Washington publication ; 635},
p. \bfpage{157}.
\bpublisher{Carnegie Institution of Washington},
\blocation{Washington, D.C.}
(\byear{1981}).
\bcomment{Allan Sandage and G.A. Tammann. ill. ; 32 cm.}
\end{bbook}
\endbibitem

\bibitem[\protect\citeauthoryear{Sandage \textit{et~al.}}{1994}]{sand94}
\begin{bbook}
\bauthor{\bsnm{Sandage}, \binits{A.}}, \bauthor{\bsnm{Bedke}, \binits{J.}},
  \bauthor{\bparticle{of~}\bsnm{Washington.}, \binits{C.I.}},
  \bauthor{\bsnm{Foundation.}, \binits{F.}}:
\bbtitle{The carnegie atlas of galaxies}.
\bsertitle{Carnegie Institution of Washington publication ; 638.},
p. \bfpage{2}.
\bpublisher{Carnegie Institution of Washington, with the Flintridge
  Foundation},
\blocation{Washington, D.C.}
(\byear{1994}).
\bcomment{by Allan Sandage and John Bedke. chiefly ill. ; 36 x 45 cm. Includes
  index (v. 2).}
\end{bbook}
\endbibitem

\bibitem[\protect\citeauthoryear{Sault, Teuben, and Wright}{1995}]{saul95}
\begin{botherref}
\oauthor{\bsnm{Sault}, \binits{R.J.}}, \oauthor{\bsnm{Teuben}, \binits{P.J.}},
  \oauthor{\bsnm{Wright}, \binits{M.C.H.}}:
1995
A retrospective view of miriad
\end{botherref}
\endbibitem

\bibitem[\protect\citeauthoryear{Schlegel, Finkbeiner, and
  Davis}{1998}]{schl98}
\begin{barticle}
\bauthor{\bsnm{Schlegel}, \binits{D.J.}}, \bauthor{\bsnm{Finkbeiner},
  \binits{D.P.}}, \bauthor{\bsnm{Davis}, \binits{M.}}:
\bjtitle{Astrophysical Journal}
\bvolume{500},
\bfpage{525}
(\byear{1998})
\end{barticle}
\endbibitem

\bibitem[\protect\citeauthoryear{Schure \textit{et~al.}}{2009}]{schu09}
\begin{barticle}
\bauthor{\bsnm{Schure}, \binits{K.M.}}, \bauthor{\bsnm{Vink}, \binits{J.}},
  \bauthor{\bsnm{Achterberg}, \binits{A.}}, \bauthor{\bsnm{Keppens},
  \binits{R.}}:
\bjtitle{Advances in Space Research}
\bvolume{44},
\bfpage{433}
(\byear{2009}).
\bcomment{DOI: 10.1016/j.asr.2009.05.010; eprintid: arXiv:0905.1134}
\end{barticle}
\endbibitem

\bibitem[\protect\citeauthoryear{Schwering and Israel}{1993}]{schw93}
\begin{botherref}
\oauthor{\bsnm{Schwering}, \binits{P.B.W.}}, \oauthor{\bsnm{Israel},
  \binits{F.P.}}:
Ir sources in magellanic clouds (schwering+ 1990).
VizieR Online Data Catalog,
vol. 2181,
p. 0
(1993).
Machine readable data available in Downloads directory
\end{botherref}
\endbibitem

\bibitem[\protect\citeauthoryear{Smith, Leiton, and Pizarro}{}]{smit00}
\begin{botherref}
\oauthor{\bsnm{Smith}, \binits{C.}}, \oauthor{\bsnm{Leiton}, \binits{R.}},
  \oauthor{\bsnm{Pizarro}, \binits{S.}}:
In: Stars, Gas and Dust in Galaxies: Exploring the Links
vol. 221,
p. 83.
ISBN: <ISBN>1-58381-053-6</ISBN>
\end{botherref}
\endbibitem

\bibitem[\protect\citeauthoryear{Sreekumar and Fichtel}{1991}]{sree91}
\begin{barticle}
\bauthor{\bsnm{Sreekumar}, \binits{P.}}, \bauthor{\bsnm{Fichtel},
  \binits{C.E.}}:
\bjtitle{Astronomy and Astrophysics}
\bvolume{251},
\bfpage{447}
(\byear{1991})
\end{barticle}
\endbibitem

\bibitem[\protect\citeauthoryear{Stanimirovic \textit{et~al.}}{1999}]{stan99}
\begin{barticle}
\bauthor{\bsnm{Stanimirovic}, \binits{S.}}, \bauthor{\bsnm{Staveley-Smith},
  \binits{L.}}, \bauthor{\bsnm{Dickey}, \binits{J.M.}}, \bauthor{\bsnm{Sault},
  \binits{R.J.}}, \bauthor{\bsnm{Snowden}, \binits{S.L.}}:
\bjtitle{Monthly Notices of the Royal Astronomical Society}
\bvolume{302},
\bfpage{417}
(\byear{1999})
\end{barticle}
\endbibitem

\bibitem[\protect\citeauthoryear{Staveley-Smith \textit{et~al.}}{1995}]{stav95}
\begin{barticle}
\bauthor{\bsnm{Staveley-Smith}, \binits{L.}}, \bauthor{\bsnm{Sault},
  \binits{R.J.}}, \bauthor{\bsnm{McConnell}, \binits{D.}},
  \bauthor{\bsnm{Kesteven}, \binits{M.J.}}, \bauthor{\bsnm{Hatzidimitriou},
  \binits{D.}}, \bauthor{\bsnm{Freeman}, \binits{K.C.}},
  \bauthor{\bsnm{Dopita}, \binits{M.A.}}:
\bjtitle{Publications of the Astronomical Society of Australia}
\bvolume{12},
\bfpage{13}
(\byear{1995})
\end{barticle}
\endbibitem

\bibitem[\protect\citeauthoryear{Staveley-Smith \textit{et~al.}}{1997}]{stav97}
\begin{barticle}
\bauthor{\bsnm{Staveley-Smith}, \binits{L.}}, \bauthor{\bsnm{Sault},
  \binits{R.J.}}, \bauthor{\bsnm{Hatzidimitriou}, \binits{D.}},
  \bauthor{\bsnm{Kesteven}, \binits{M.J.}}, \bauthor{\bsnm{McConnell},
  \binits{D.}}:
\bjtitle{Monthly Notices of the Royal Astronomical Society}
\bvolume{289},
\bfpage{225}
(\byear{1997})
\end{barticle}
\endbibitem

\bibitem[\protect\citeauthoryear{Voelk}{1989}]{voel89}
\begin{barticle}
\bauthor{\bsnm{Voelk}, \binits{H.J.}}:
\bjtitle{Astronomy and Astrophysics}
\bvolume{218},
\bfpage{67}
(\byear{1989})
\end{barticle}
\endbibitem

\bibitem[\protect\citeauthoryear{Weingartner and Draine}{2001}]{wein01}
\begin{barticle}
\bauthor{\bsnm{Weingartner}, \binits{J.C.}}, \bauthor{\bsnm{Draine},
  \binits{B.T.}}:
\bjtitle{Astrophysical Journal}
\bvolume{548},
\bfpage{296}
(\byear{2001})
\end{barticle}
\endbibitem

\bibitem[\protect\citeauthoryear{Westerlund}{1997}]{west97}
\begin{bbook}
\bauthor{\bsnm{Westerlund}, \binits{B.E.}}:
\bbtitle{The magellanic clouds}.
\bsertitle{Cambridge astrophysics series ; 29},
p. \bfpage{279}.
\bpublisher{Cambridge University Press},
\blocation{New York}
(\byear{1997}).
\bcomment{Bengt E. Westerlund. ill.}
\end{bbook}
\endbibitem

\bibitem[\protect\citeauthoryear{Wong \textit{et~al.}}{2011}]{wong11}
\begin{barticle}
\bauthor{\bsnm{Wong}, \binits{G.F.}}, \bauthor{\bsnm{Filipovic},
  \binits{M.D.}}, \bauthor{\bsnm{Crawford}, \binits{E.J.}},
  \bauthor{\bparticle{de~}\bsnm{Horta}, \binits{A.Y.}}, \bauthor{\bsnm{Galvin},
  \binits{T.}}, \bauthor{\bsnm{Draskovic}, \binits{D.}}, \bauthor{\bsnm{Payne},
  \binits{J.L.}}:
\bjtitle{Serbian Astronomical Journal}
\bvolume{182},
\bfpage{43}
(\byear{2011})
\end{barticle}
\endbibitem

\bibitem[\protect\citeauthoryear{Wong \textit{et~al.}}{2011a}]{wong11a}
\begin{barticle}
\bauthor{\bsnm{Wong}, \binits{G.F.}}, \bauthor{\bsnm{Filipovic},
  \binits{M.D.}}, \bauthor{\bsnm{Crawford}, \binits{E.J.}},
  \bauthor{\bsnm{Tothill}, \binits{N.F.H.}},
  \bauthor{\bparticle{de~}\bsnm{Horta}, \binits{A.Y.}},
  \bauthor{\bsnm{Draskovic}, \binits{D.}}, \bauthor{\bsnm{Galvin},
  \binits{T.J.}}, \bauthor{\bsnm{Collier}, \binits{J.D.}},
  \bauthor{\bsnm{Payne}, \binits{J.L.}}:
\bjtitle{Serbian Astronomical Journal}
\bvolume{183},
\bfpage{103}
(\byear{2011}a)
\end{barticle}
\endbibitem

\bibitem[\protect\citeauthoryear{Wong \textit{et~al.}}{2012}]{wong12}
\begin{barticle}
\bauthor{\bsnm{Wong}, \binits{G.F.}}, \bauthor{\bsnm{Crawford}, \binits{E.J.}},
  \bauthor{\bsnm{Filipovic}, \binits{M.D.}}, \bauthor{\bsnm{De~Horta},
  \binits{A.Y.}}, \bauthor{\bsnm{Tothill}, \binits{N.F.H.}},
  \bauthor{\bsnm{Collier}, \binits{J.D.}}, \bauthor{\bsnm{Draskovic},
  \binits{D.}}, \bauthor{\bsnm{Galvin}, \binits{T.J.}}, \bauthor{\bsnm{Payne},
  \binits{J.L.}}:
\bjtitle{ArXiv e-prints}
\bvolume{1203},
\bfpage{4310}
(\byear{2012}).
\bcomment{25 pages, 2 tables, submitted to SAJ}
\end{barticle}
\endbibitem

\bibitem[\protect\citeauthoryear{Xu \textit{et~al.}}{1992}]{xu92}
\begin{barticle}
\bauthor{\bsnm{Xu}, \binits{C.}}, \bauthor{\bsnm{Klein}, \binits{U.}},
  \bauthor{\bsnm{Meinert}, \binits{D.}}, \bauthor{\bsnm{Wielebinski},
  \binits{R.}}, \bauthor{\bsnm{Haynes}, \binits{R.F.}}:
\bjtitle{Astronomy and Astrophysics}
\bvolume{257},
\bfpage{47}
(\byear{1992})
\end{barticle}
\endbibitem

\bibitem[\protect\citeauthoryear{Ye and Turtle}{1991}]{ye91}
\begin{barticle}
\bauthor{\bsnm{Ye}, \binits{T.}}, \bauthor{\bsnm{Turtle}, \binits{A.J.}}:
\bjtitle{Monthly Notices of the Royal Astronomical Society}
\bvolume{249},
\bfpage{693}
(\byear{1991})
\end{barticle}
\endbibitem

\bibitem[\protect\citeauthoryear{Yun, Reddy, and Condon}{2001}]{yun01}
\begin{barticle}
\bauthor{\bsnm{Yun}, \binits{M.S.}}, \bauthor{\bsnm{Reddy}, \binits{N.A.}},
  \bauthor{\bsnm{Condon}, \binits{J.J.}}:
\bjtitle{Astrophysical Journal}
\bvolume{554},
\bfpage{803}
(\byear{2001})
\end{barticle}
\endbibitem

\bibitem[\protect\citeauthoryear{Zaritsky \textit{et~al.}}{2000}]{zari00}
\begin{barticle}
\bauthor{\bsnm{Zaritsky}, \binits{D.}}, \bauthor{\bsnm{Harris}, \binits{J.}},
  \bauthor{\bsnm{Grebel}, \binits{E.K.}}, \bauthor{\bsnm{Thompson},
  \binits{I.B.}}:
\bjtitle{Astrophysical Journal}
\bvolume{534},
\bfpage{53}
(\byear{2000})
\end{barticle}
\endbibitem

\end{thebibliography}

\end{document}